\documentclass[reprint,prd,onecolumn,notitlepage,11pt]{revtex4-1}
\usepackage[english]{babel}
\usepackage{amsmath,amssymb,graphicx,bm,lipsum}
\usepackage[colorlinks=true,citecolor=blue,linkcolor=blue,urlcolor=blue]{hyperref} 
\usepackage[font={small},flushleft,indent]{caption}

\begin{document}

\title{Aberration effect on lower-order images of thin accretion disk in the astrometric approach}
\author{Qing-Hua Zhu}
\email{zhuqh@cqu.edu.cn}
\affiliation{Department of Physics, Chongqing University, Chongqing 401331, China}

\begin{abstract}
  With recent advancements in observing supermassive black holes with the Event Horizon Telescope, there has been persistent exploration into what the images can reveal about fundamental physics, including space-time geometries and astrophysical emission sources. Inspired by Penrose's aberration formula for a rigid sphere, which clarified that increased speed does not flatten the appearance of the sphere, we extend the studies to the behavior of the images of accretion emissions. This paper examines the impact of aberration effects on the images of a thin accretion disk around Kerr-de Sitter black holes for finite distant observers, specifically focusing on the primary, secondary, and $n=2$ images. We employ the analytical ray-tracing scenario and extend the astrometric approach to investigate the images in the presence of aberration. This study is non-trivial because we do not assume a specific form of the aberration formula, instead, all aberration effects emerge from a coordinate-independent and tetrad-independent framework referred to as the astrometric approach. Our study finds that the shapes of the lower-order images get highly distorted for finite observers in motion, and the shapes and sizes of primary images are more sensitive to aberration than those of the $n=2$ images. This finding suggests that the primary images could theoretically be distinguished from the shadow based on their distinctive variations. 
\end{abstract} 

\maketitle

\

\

\section{introduction}

Since the images of the supermassive black holes at horizon scale were taken by the Event Horizon Telescope (EHT) \cite{EventHorizonTelescope:2019dse, EventHorizonTelescope:2019ggy, EventHorizonTelescope:2019pgp,  EventHorizonTelescope:2022xnr}, what the images can reveal about fundamental physics, including space-time geometries, or astrophysical emissions sources, has been persistently explored \cite{Narayan:2019imo, Gralla:2019xty, Mizuno:2018lxz,  EventHorizonTelescope:2020qrl, Gralla:2020pra, Bronzwaer:2021lzo, Vincent:2022fwj}.

In the vicinity of a black hole, the gravitational bending of light could give rise to the multiple imaging of emission sources, which are known as primary, secondary, and generally $n$th-order images \cite{Virbhadra:1999nm, Virbhadra:2008ws, Bisnovatyi-Kogan:2022ujt}. 
The higher-order images of the emissions, which also result in the boundary of black hole shadow \cite{Falcke:1999pj}, were shown to be insensitive to emission sources, and thus can theoretically be utilized to reflect the gravity of a black hole \cite{Bardeen:1973tla, Falcke:1999pj, Narayan:2019imo}. 
In this context, the shadows of various physically motivated black holes were studied, with the aim of establishing a connection between black hole parameters and EHT observations \cite{EventHorizonTelescope:2021dqv, Tan:2023ngk, Afrin:2023uzo, Olmo:2023lil}, or exploring the possibility of testing Einstein's theory of gravity \cite{Mizuno:2018lxz, EventHorizonTelescope:2020qrl}.
However, due to the current precision of the EHT, the shadow may not be visible in the images \cite{Johnson:2019ljv, Gralla:2020pra}. The black hole images, captured through primary or secondary emissions, might suggest that considering the properties of the emission sources is inevitable \cite{Gralla:2019xty}. As known, the shadow corresponds to the center dark region of the images for spherical accretion emission \cite{Falcke:1999pj, Narayan:2019imo, Bronzwaer:2021lzo}, while it is not the case when considering accretion disk emissions \cite{Gralla:2019xty, Johnson:2019ljv, Vincent:2022fwj}. The investigation into the behaviors of lower-order images for given emissions, even in the cases of non-Kerr black holes, appears to be well-motivated \cite{Gan:2021pwu, daSilva:2023jxa, Wang:2023vcv, Cimdiker:2023zdi, Huang:2023ilm, Gjorgjieski:2023qpv, Rosa:2023hfm, Roder:2023oqa, Younsi:2021dxe}.

To study the lower-order images, there could be an influence from the aberration of light. Half a century ago, Penrose derived the aberration formulae for a rigid sphere \cite{1959PCPS...55..137P}, indicating that the appearance of the sphere would not appear to be flattened as the speed increases. Recently, similar results have been found for the black hole shadow in the view of observers in motion \cite{Chang:2020lmg, Chang:2021ngy}. The difference is that, in the presence of gravity, the shape of the shadow was shown to be observer-dependent. Hence, it is reasonable to anticipate that information about space-time geometries can be gleaned from the aberration effect. While the aberration effect might be considered ignorable because there seem to be no physical mechanisms that can provide such high speeds, recent studies have pointed out that observers co-moving with the expansion of the universe at spatial infinity can still observe a finite-size shadow \cite{Perlick:2018iye, Chang:2019vni, Li:2020drn}. In this case, the aberration effect can not be ignored because the size of the shadow increases with co-moving speeds, offsetting the decrease in distance.

In this paper, we investigate the behaviors of primary, secondary, and \(n=2\) images of the thin accretion disk around a Kerr-de Sitter black hole, considering the aberration effect for finite distant observers. In addition to co-moving observers, we explore the variations of the images concerning observers in radial and axial motions.
On the technical aspects, we present an alternative form of the transfer functions in the analytical ray-tracing scenario developed by pioneers \cite{Gralla:2019ceu, Gralla:2019drh, Cardenas-Avendano:2022csp}. We also extend the astrometric approach to establish observers' celestial sphere \cite{Chang:2020miq, Chang:2021ngy}. Notably, our approach avoids assuming a specific form of the aberration formula, instead, all aberration effects emerge from the astrometric approach. As anticipated, our study shows that the shapes of lower-order images become distorted in the view of observers in motion. The variations in image sizes align with Penrose's aberration formulae overall. 
Our results indicate that the variation of primary and secondary images is more sensitive to observers' motion than \(n=2\) images. This finding suggests that primary or even secondary images could theoretically be distinguished from the shadow.

The rest of the paper is organized as follows. In Section~\ref{II}, we present the geodesic equations of light and evaluate them using hyperbolic and trigonometric coordinates to derive the transfer function. Section~\ref{III} focuses on establishing the observers' celestial sphere through the utilization and extension of the astrometric approach. In Section~\ref{IV}, we simplify the transfer functions, taking into account the thin accretion disk. The qualitative and quantitative results of the aberration effect on the primary, secondary, and \(n=2\) images of the thin accretion disk are presented in Section~\ref{V}. Finally, Section~\ref{VI} provides a summary of conclusions and discussions.

\section{Ray tracing equations \label{II}}

In this section, we follow the analytical ray-tracing scenario developed by pioneers \cite{Gralla:2019ceu, Gralla:2019drh, Cardenas-Avendano:2022csp}. To address turning points on the trajectories of light, we employ hyperbolic and trigonometric coordinates as an alternative for evaluating the transfer functions. As mentioned in the introduction, our focus is on the images of emissions concerning both finite distant observers and co-moving observers. Thus the Kerr-de Sitter black hole is considered in this context.

\subsection{Geodesic equations}
\label{II.A}

Kerr de-Sitter metric can describe a rotating black hole with mass $M$ and spin $a$ in the presence of cosmological constant $\Lambda$, which is given by
\begin{eqnarray}
  {\rm d} s^2 & = & - \frac{\Delta_r}{\Sigma} ({\rm d} t - {\rm asin }^2 \theta
  {\rm d} \phi)^2 + \frac{\Delta_{\theta} \sin^2 \theta}{\Sigma} (a {\rm d} t -
  (r^2 + a^2) {\rm d} \phi)^2 + \frac{\Sigma}{\Delta_r} {\rm d} r^2 +
  \frac{\Sigma}{\Delta_{\theta}} {\rm d} \theta^2~, \label{met}
\end{eqnarray}
where
\begin{subequations}
  \begin{eqnarray}
    \Delta_r (r) & = & - \frac{1}{3} \Lambda r^2 (r^2 + a^2) + r^2 - 2 M
    r + a^2~,\\
    \Delta_{\theta} (\theta) & = & 1 + \frac{1}{3} \Lambda a^2 \cos^2 \theta~,\\
    \Sigma & = & r^2 + a^2 \cos^2 \theta~.
  \end{eqnarray}
\end{subequations}
For the stationary black hole, the event horizon can be determined by the equation $\Delta_r(r)=0$. Besides the horizon at $r \simeq M$, there is an outer horizon due to the cosmological constant $\Lambda$. In co-moving coordinates, the outer horizon could be understood as the cosmological horizon.
Additionally, there could be alternative coordinates of time $\tilde{t}$ and azimuth angle $\tilde{\phi}$ defined by $\tilde{t} = \Xi t$ and $\tilde{\phi} = \Xi \phi$, where $\Xi = 1 + \frac{1}{3} \Lambda a^2$ \cite{Grenzebach:2014fha}. We will clarify later that the coordinate choice does not affect the results in this paper.

Employing the Hamilton-Jacobi method for geodesic equations, the action can be obtained in the form of
\begin{eqnarray}
  \mathcal{S}  =  - E   t + L \phi \pm_r \int {\rm d} r \left\{
  \frac{\sqrt{((r^2 + a^2) E - a   L)^2 - \Delta_r K}}{\Delta_r}
  \right\} \pm_\theta \int {\rm d} \theta \left\{ \frac{1}{\Delta_{\theta}}
  \sqrt{\Delta_{\theta} K - \frac{(a   {\rm{Esin}}^2 \theta -  L)^2}{\sin^2 \theta}} \right\}~,\nonumber\\
\end{eqnarray}
where $E$, $L$, and $K$ are integral constants due to the intrinsic symmetry of the space-time. Using the action $\mathcal{S}$, one can derive the 4-momentum of light with $p_\mu=\partial{\mathcal{S}}/\partial{x^\mu}$, which leads to
\begin{subequations}
  \begin{eqnarray}
    p_t & = & - E ~,\\
    (p_r)^2 & = & \frac{(E (r^2 + a^2) - a   L)^2 - \Delta_r
      K}{\Delta_r^2} ~,\\
    (p_{\theta})^2 & = & \frac{1}{\Delta_{\theta}^2} \left( \Delta_{\theta} K -
    \frac{(L - E  a {\sin }^2 \theta)^2}{\sin^2 \theta} \right) ~,\\
    p_{\phi} & = & L ~.
  \end{eqnarray}\label{mom:p}
\end{subequations}
Because $E$, $K$, and $L$ are constants in $\mathcal{S}$, namely, $\partial
  \mathcal{S}/ \partial E = \partial \mathcal{S}/ \partial K = \partial
  \mathcal{S}/ \partial L = 0$, we can rewrite the geodesic equations in terms of  integrals, namely,
\begin{subequations}
  \begin{eqnarray}
    0 & = & - t +\mathcal{I}_t + a^2 \left( 1 + \frac{1}{3} \Lambda (a^2 - a
      \lambda) \right) \mathcal{G}_t \text{} ~,\label{geo1}\\
    0 & = & - \phi +\mathcal{I}_{\phi} + \lambda \mathcal{G}_{\phi} +
    \frac{1}{3} \Lambda a^3 \mathcal{G}_t ~,\label{geo2}\\
    0 & = & \mathcal{G}_{\theta} -\mathcal{I}_r  \label{geo3}~,
  \end{eqnarray} \label{geo:int}
\end{subequations}
where the indefinite integrals are given by
\begin{subequations}
  \begin{eqnarray}
    \mathcal{I}_r (r) & \equiv & \pm_r\int \frac{{\rm d} r}{\sqrt{\mathcal{R} (r)}} \label{int1}~,\\
    \mathcal{I}_t (r) & \equiv & \pm_r\int {\rm d} r \left\{ \frac{r^2 \Delta_r + \left( \frac{1}{3} \Lambda r^2 (r^2 + a^2) + 2 M   r \right) (r^2 + a^2 - a \lambda)}{\Delta_r \sqrt{\mathcal{R} (r)}} \right\} ~,\\
    \mathcal{I}_{\phi} (r) & \equiv & \pm_r\int {\rm d} r \left\{ \frac{a \left( 2 M r - a \lambda - \frac{1}{3} \Lambda r^2 (r^2 + a^2)
      \right)}{\Delta_r \sqrt{\mathcal{R} (r)}} \right\} ~,\\
    \mathcal{G}_{\theta} (\theta) & \equiv & \pm_\theta\int \frac{{\rm d} \theta}{\sqrt{\Theta (\theta)}}  \overset{}{} \label{int4}~,\\
    \mathcal{G}_t (\theta) & \equiv & \pm_\theta\int {\rm d} \theta \left\{\frac{  \cos^2 \theta  }{\Delta_{\theta} \sqrt{\Theta
        (\theta)}} \right\} ~,\\
    \mathcal{G}_{\phi} (\theta) & \equiv & \pm_\theta \int {\rm d} \theta \left\{\frac{\csc^2 \theta}{\Delta_{\theta} \sqrt{\Theta (\theta)}} \right\} ~.
  \end{eqnarray} \label{geo:int:where}
\end{subequations}
In above equaions, we have let quantities $\lambda \equiv L / E$ and $\kappa \equiv
  K / E^2$, and
\begin{subequations}
  \begin{eqnarray}
    \mathcal{R} (r) & \equiv & (r^2 + a^2 - a   \lambda)^2 - \Delta_r
    \kappa~,\\
    \Theta (r) & \equiv & \Delta_{\theta} \kappa - \frac{(\lambda - a
      \sin^2 \theta)^2}{\sin^2 \theta}~.
  \end{eqnarray}
\end{subequations}
It should be noted that $\pm_r$ and $\pm_\theta$ in Eqs~(\ref{geo:int:where}) are not constant. They transition from $\pm$ to $\mp$ when a light ray encounters a turning point in propagation. Specifically, $\pm_r$ changes when the light ray reaches its minimum distance to the black hole, and $\pm_\theta$ changes when the light ray completes half an orbit.

\subsection{Transfer functions }
\label{II.B}

From the emission regions to the observers (image plane), one can employ Mino time $\tau$ for evaluating the geodesic equations,
\begin{eqnarray}
  \tau & \equiv & G_{\theta} (\theta_{\rm o}, \theta_{\rm s}) =\mathcal{G}_{\theta}
  (\theta_{\rm o}) -\mathcal{G}_{\theta} (\theta_{\rm s}) \text{}  \label{raytrace1}~, \label{raytrace1}
\end{eqnarray}
where $\theta_{\rm o}$ and $\theta_{\rm s}$ denote the inclination angles of  observer and emission source. The Mino time $\tau$ is monotonic as the propagation of light, whereas the polar angle $\theta$ might not. 
Based on Eqs.~(\ref{geo3}) and (\ref{raytrace1}), radial location of emission $r_{\rm s}$ can
be reversely ray-traced from the location of observers $r_{\rm o}$, namely,
\begin{eqnarray}
  r_{\rm s} & = & I_r^{- 1} (r_{\rm o} ; \tau) ~, \label{raytrace2}
\end{eqnarray}
where $ I_r (r_{\rm o}, r_{\rm s}) \equiv \mathcal{I}_r
  (r_{\rm s}) -\mathcal{I}_r (r_{\rm o})$. Based on Eqs.~(\ref{geo1}) and (\ref{geo2}), one can obtain azimuth angle $\phi_{\rm s}$ and time $t_{\rm s}$ at the emission regions as follows
\begin{eqnarray}
  \phi_{\rm s} & = & \phi_{\rm o} - I_{\phi} (r_{\rm o} , r_{\rm s}) - \lambda G_{\phi} (\theta_{\rm o},
  \theta_{\rm s}) - \frac{\Lambda}{3} a^3 G_t (\theta_{\rm o}, \theta_{\rm s}) ~,\label{raytrace3}\\
  t_{\rm s} & = & t_{\rm o} - I_t (r_{\rm o} , r_{\rm s}) - a^2 \left( 1 + \frac{\Lambda}{3} (a^2 - a
    \lambda) \right) G_t (\theta_{\rm o}, \theta_{\rm s})  \label{raytrace4}~,
\end{eqnarray}
where $\phi_{\rm o}$ and $t_{\rm o}$ are azimuth angle and time of observers, respectively, $G_{\ast} (\theta_{\rm o}, \theta_{\rm s}) \equiv \mathcal{G}_{\ast} (\theta_{\rm o})
  -\mathcal{G}_{\ast} (\theta_{\rm s})$ and $I_{\ast} (\theta_{\rm o}, \theta_{\rm s}) \equiv
  \mathcal{I}_{\ast} (\theta_{\rm o}) -\mathcal{I}_{\ast} (\theta_{\rm s})$.
  Eqs.(\ref{raytrace2})--(\ref{raytrace4}) are also referred to as transfer functions \cite{Cardenas-Avendano:2022csp}. These functions can map the location of observers $(r_{\rm o}, \theta_{\rm o}, \phi_{\rm o})$ to the emission regions $(r_{\rm s}, \theta_{\rm s}, \phi_{\rm s})$. Unlike the transfer functions for Kerr black holes \cite{Gralla:2019ceu, Gralla:2019drh, Cardenas-Avendano:2022csp}, it is found that Eqs.(\ref{raytrace3}) and (\ref{raytrace4}) include an additional angular integral $G_t (\theta_{\rm o}, \theta_{\rm s})$ due to the cosmological constant.

To conduct a reverse ray-tracing procedure based on Eqs.~(\ref{raytrace1})--(\ref{raytrace4}), we have following steps,
\begin{itemize}
  \item[i.] Compute the Mino time $\tau$ for given polar angles $\theta_{\rm o}$, $\theta_{\rm s}$, and the number of half orbits $n$.
  \item[ii.] Obtain $r_{\rm s}$ for given radial location of observers $r_{\rm o}$ with the known Mino time $\tau$.
  \item[iii.] Compute the azimuth angle $\phi_{\rm s}$ and time $t_{\rm s}$ of emissions for given $\phi_{\rm o}$ and $t_{\rm o}$ with the known polar angles $(\theta_{\rm o}, \theta_{\rm s})$ and radial locations $(r_{\rm o}, r_{\rm s})$.
\end{itemize}
In the following, we present explicit expressions of Eq.~(\ref{raytrace1}) and (\ref{raytrace2}) for studying axisymmetric emission models. To address the turning points mentioned in Section~{\ref{II.A}}, we alternatively utilize hyperbolic and trigonometric coordinates for evaluating the transfer functions.

\

\subsubsection{Computation of $G_{\theta} (\theta_{\rm o}, \theta_{\rm s})$}

Based on Eqs.~(\ref{int4}) and (\ref{raytrace1}), the integral
$G_{\theta}$ here can be rewritten as
\begin{eqnarray}
  G_{\theta} (\theta_{\rm o}, \theta_{\rm s}) & = & \pm_{\theta}
  \int^{\theta_{\rm o}}_{\theta_{\rm s}} \frac{{\rm d} \theta}{\sqrt{\mathbb{A}+\mathbb{B}
      \cot^2 \theta + \mathbb{C} \cos^2 \theta}}  \label{Ev1}~,
\end{eqnarray}
where $\mathbb{A}= \kappa-(\lambda-a)^2$, $\mathbb{B}= - \lambda^2$, and ${\mathbb{C}}= a^2 (1 +  \Lambda \kappa/ 3) $. The $\pm_{\theta}$ changes into $\mp_{\theta}$ when the light encounters tuning points.
To deal with the turning points, we introduce variable $\chi$ such that
\begin{eqnarray}
  \cos \theta & = & \sqrt{u_+} \cos \chi  \label{sub1}~,
\end{eqnarray}
where the $u_\pm$ are given by the roots of the equation $\Theta (\theta)=0$, namely,
\begin{eqnarray}
  u_{\pm} & \equiv & \cos^2 \theta_{\pm} =
  \frac{{\mathbb{C}}+\mathbb{B}-\mathbb{A} \pm
    \sqrt{({\mathbb{C}}+\mathbb{B}-\mathbb{A})^2 -
      4\mathbb{A}{\mathbb{C}}}}{2{\mathbb{C}}}~.
\end{eqnarray}
For a given null geodesic with fixed $\kappa$ and $\lambda$, the polar angle $\theta$ on the geodesic varies between $-\sqrt{u_+}$ and $\sqrt{u_+}$, while the variable $\chi$ can monotonically increase or decrease.
By making use of the above variable substitution, the integral in
Eq.~(\ref{Ev1}) reduces to
\begin{eqnarray}
  G_{\theta} & = & \pm \int^{\chi_{\rm o}}_{\chi_{\rm s}} \frac{{\rm d} \chi}{\sqrt{\mathbb{C}}
    \sqrt{u_+ \cos^2 \chi - u_-}} = \left.\pm \frac{1}{\sqrt{\mathbb{C}}} \frac{1}{\sqrt{u_+ -
      u_-}} F \left( \chi, \frac{u_+}{u_+ - u_-} \right) \right|^{\chi_{\rm o}}_{\chi_{\rm s}}~,
\end{eqnarray}
where $\pm$ is constant, and $F (a, b)$ is the Bessel function of the first kind. By
inversely solving Eq.~(\ref{sub1}), we can obtain
\begin{subequations}
  \begin{eqnarray}
    \chi_{\rm o} & = & \arccos \left( \frac{1}{\sqrt{u_+}} \cos \theta_{\rm o} \right)~, \label{chi:o}\\
    \chi_{\rm s} & = & \left( m + \frac{1 - (- 1)^m}{2} \right) \pi + (- 1)^m \arccos
    \left( \frac{1}{\sqrt{u_+}} \cos \theta_{\rm s} \right) ~. \label{chi:s}
  \end{eqnarray}
\end{subequations}
Here, we adopt the convention $\chi_{\rm o} \in [0, \pi]$. The integer $m$ can be related to the number of half-orbits with respect to the initial locations.

\

\subsubsection{Computation of $I_r(r_{\rm o},r_{\rm s})$}

From Eq.~(\ref{int1}), the integral $I_r$ can be rewritten as
\begin{eqnarray}
  I_r (r_{\rm o}, r_{\rm s}) & = & \pm_r \int^{r_{\rm s}}_{r_{\rm s}} \frac{{\rm d} r}{\sqrt{C (r^4
      +\mathcal{A}r^2 +\mathcal{B}   r +\mathcal{C})}}  \label{Ev2}~,
\end{eqnarray}
where $C=a^{-2}\mathbb{C}$, $\mathcal{A}= a^2 C -\kappa -2a \lambda+a^2$, $\mathcal{B}= 2 \kappa  M$, $\mathcal{C}= - a^2 (\kappa-(\lambda-a)^2)$. Similar to $\pm_{\theta}$, the $\pm_{r}$ transitions into $\mp_{r}$ when the light encounters tuning points.  To address the turning points of $\mathcal{R} (r)$ at $r = r_4$, we introduce variable $\xi$ such that
\begin{eqnarray}
  r & = & r_4 \cosh^2 \xi  \label{sub2}~.
\end{eqnarray}
where $r_4$ is given by the roots of equation $\mathcal{R} (r) = 0$, namely,
\begin{eqnarray}
  r_{1, 2, 3, 4} & = & \pm^{(\ast)} \frac{1}{2} \sqrt{\mathfrak{g}-
    \frac{2\mathcal{A}}{3}} \pm^{(\star)} \frac{1}{2} \sqrt{\mp^{(\ast)}
    \frac{2\mathcal{B}}{\sqrt{\mathfrak{g}- \frac{2\mathcal{A}}{3}}} -
    \frac{4\mathcal{A}}{3} -\mathfrak{g}} ~,
\end{eqnarray}
the $\pm^{(\ast)}$ and $\pm^{(\star)}$ are independent, thereby resulting in four distinct roots denoted by the subscripts 1,2,3,4, and
\begin{subequations}
  \begin{eqnarray}
    \mathfrak{g} & = & \frac{1}{3} \left( 2^{\frac{1}{3}} \left(
    \frac{\mathfrak{q}}{\mathfrak{w}} \right) + 2^{- \frac{1}{3}} \mathfrak{w}
    \right) ~,\\
    \mathfrak{w} & = & \left( 2\mathcal{A}^3 + 27\mathcal{B}^2 -
    72\mathcal{A}\mathcal{C}+ \sqrt{- 4\mathfrak{q}^3 + (2\mathcal{A}^3 +
      27\mathcal{B}^2 - 72\mathcal{A}\mathcal{C})^2} \right)^{\frac{1}{3}} ~,\\
    \mathfrak{q} & = & \mathcal{A}^2 + 12\mathcal{C} ~.
  \end{eqnarray}
\end{subequations}
We let $r_4$ and $r_3$ represent the two largest real roots, and $r_4
  \geqslant r_3$.
By making use of the variable substitution, the integral in Eq.~(\ref{Ev2}) reduces to
\begin{eqnarray}
  I_r & = & \pm \frac{2}{r_4 \sqrt{C}} \int^{\xi_{\rm o}}_{\xi_{\rm s}} \frac{\cosh \xi
    {\rm d} \xi}{\sqrt{\left( \cosh^2 \xi - \frac{r_3}{r_4} \right) \left(
      \cosh^2 \xi - \frac{r_2}{r_4} \right) \left( \cosh^2 \xi - \frac{r_2}{r_4}
      \right)}} \nonumber\\
  & = & \left. \pm \frac{2}{\sqrt{C} \sqrt{r_{31} r_{42}}} F \left( \arcsin
  \left( \frac{\sinh \xi}{\sqrt{\frac{r_{41}}{r_{31}} \left( \cosh^2 \xi -
      \frac{r_3}{r_4} \right)}} \right), \frac{r_{32} r_{41}}{r_{42} r_{31}}
  \right) \right|^{\xi_{\rm o}}_{\xi_{\rm s}} ~, \label{Ir}
\end{eqnarray}
where $\pm$ is constant, and $r_{i   j} \equiv r_i - r_j$.
By inversely solving Eq.~(\ref{sub2}), we obtain
\begin{eqnarray}
  \xi_{\ast} & = & \pm {\rm arccosh} \left( \sqrt{\frac{r_{\ast}}{r_4}}
  \right), \hspace{1.5em} \text{for } \ast = \rm o, s~. \label{xi:os}
\end{eqnarray}
The choice of $\pm$ in Eq.~(\ref{xi:os}) can represent different situations. For example, there would be a turning point, if $\xi_{\rm s} \xi_{\rm o} <
  0$.

Finally, we obtain the transfer function in Eq.~(\ref{raytrace2}) based on the inverse function of Eq.~(\ref{Ir}), namely,
\begin{eqnarray}
  r_{\rm s} & = & r_3 + \frac{r_4 - r_3}{1 - \frac{r_{41}}{r_{31}} {\rm sn}^2
  \left( \pm \frac{\sqrt{C   r_{31} r_{32}}}{2} (\mathcal{I}_r (\xi_{\rm o}) - I_r (\xi_{\rm o}, \xi_{\rm s})), \frac{r_{32} r_{41}}{r_{42} r_{31}} \right)}~.
\end{eqnarray}
where $\text{sn}(a,b)$ is Jacobi elliptic function.

\

\section{Transfer functions for thin accretion disk \label{III}}

To investigate the influence of light aberration on the lower-order images of emissions, we employ an ideal and phenomenological emission model, specifically the thin accretion disk. For both optically thin and geometrically thin accretion flows, the observed intensity $I_{{\rm obs}} ({\rm image})$ on the image plane is determined by the accumulation of redshifted emission intensity  \cite{Jaroszynski:1997bw},
\begin{eqnarray}
  I_{{ \rm obs}} ({\rm image}) & = & \sum_{\gamma} (1 + z (\bm x_{\rm s}(\gamma)))^3
  I_{{ \rm emt}} \left( \bm{x}_{\rm s} (\gamma) \right) ~.
\end{eqnarray}
where $\gamma$ is the trajectories of light rays within emission regions and the
redshit can be given by
\begin{eqnarray}
  1 + z \left( \bm{x}_{\rm s} \right) & = & \frac{u_{\mu} p^{\mu}
  |_{\bm{x}_{\rm o}}}{u_{\mu} p|_{\bm{x}_{\rm s}}} ~. \label{red}
\end{eqnarray}
For the given image plane, the observer's location $\bm{x}_{\rm o}$ is fixed at the moment, and the source location ${\bm x}_{\rm s}$ is throughout the entire emission regions. The configuration of the image plane for observers situated at a finite distance will be introduced in the subsequent section. The distributions of emission intensity for the thin accretion disk
$I_{{ \rm emt}} \left( \bm{x} \right)$ can be derived from
a specific physical emission model \cite{EventHorizonTelescope:2019pgp,Johnson:2019ljv,Cardenas-Avendano:2022csp}. Here, we consider a phenomenological form of the intensity, namely,
\begin{eqnarray}
  I_{{ \rm emt}} \left( \bm{x} \right) & = & \left\{\begin{array}{ll}
    f_{\text{d}} (r) \Theta \left( r_{\text{d,+}} - r \right) \Theta \left( r
    - r_{\text{d,} -} \right) & \theta = \frac{\pi}{2}    \\
    0                         & \theta \neq \frac{\pi}{2}
  \end{array}\right. ~, \label{Iemt}
\end{eqnarray}
where $f (r)$ is the distribution of emission intensity and the width of the
disk is $r_{\text{d,+}} - r_{\text{d,} -}$. 

Because of the axisymmetries in the thin disk emission $I_{{ \rm emt}} \left( \bm{x}_{\rm s} \right)
  = I_{{ \rm emt}} (r_{\rm s}, \theta_{\rm s})$, one can employ the ray-tracing procedure presented in Sec.~\ref{II.B} with the essential equations summarized as follows:
\begin{subequations}
  \begin{eqnarray}
    \tau & = & \frac{1}{a \sqrt{C (u_+ - u_-)}} F \left. \left( \chi,
    \frac{u_+}{u_+ - u_-} \right) \right|^{\chi_{\rm o}}_{\chi_{\rm s}} ~, \\
    r_{\rm s} & = & r_3 + \frac{r_4 - r_3}{1 - \frac{r_{41}}{r_{31}} {\rm sn}^2
    \left( \pm \frac{\sqrt{C   r_{31} r_{32}}}{2} (\mathcal{I}_r
    (\xi_{\rm o}) - \tau), \frac{r_{32} r_{41}}{r_{42} r_{31}} \right)} ~,\\
    \theta_{\rm s} & = & \arccos \left( \sqrt{u_+} \cos \chi_{\rm s} \right)~,
  \end{eqnarray}\label{raytrace:disk}
\end{subequations}
where the $\mathcal{I}_r$, $\xi_{\rm o}$, $u_\pm$, $r_i$, $r_{ij}$, $C$ and  $\chi_{\rm o}$ have been defined in Sec.~\ref{II.B}, and $\chi_{\rm s}$ can be further simplified by substituting $\theta_{\rm s}=\pi/2$ in Eq.~(\ref{chi:s}), namely,
\begin{eqnarray}
  \chi_{\rm s} & = & \pi \left( m + \frac{1}{2} \right)~,
\end{eqnarray}
where the $m$ is an integer that can determine the order of images. In Table~\ref{T1}, we provide instances to illustrate the relations among $m$, $\chi_{\rm s}$, and the $n$th-order images. Since there is no relevance with the time and azimuth angle in the transfer functions, the renormalized coordinate $\tilde{t}$ and $\tilde{\phi}$ can not affect our results.
\begin{table}
  \centering
  \caption{The values of $\chi_{\rm s}$ and $m$ for $n$-order images.}\label{T1}
  \begin{ruledtabular}
    \begin{tabular}{c|cc}
      \text{images}                  & $|\chi_{\rm s}|$  & $m$            \\
      \colrule
      \text{primary (direct)}, $n=0$ & $\pi / 2$         & $0, - 1$       \\
      \text{secondary}, $n=1$        & $3\pi / 2$        & $1, - 2$       \\
      ...                            & ...               & ...            \\
      $n$\text{th-order}, $n\geq0$   & $\pi (n + 1 / 2)$ & $n$, $- n - 1$ \\
    \end{tabular}
  \end{ruledtabular}
\end{table}

By utilizing variables $\chi$ and $\xi$, the transfer functions presented in Eqs.~(\ref{raytrace:disk}) may offer a more concise form than those in pioneering studies \cite{Gralla:2019ceu, Gralla:2019drh, Cardenas-Avendano:2022csp}. However, its advantage in the ray-tracing procedure has yet to be demonstrated.

\

\section{Observers' Celestial Sphere with astrometric observables \label{IV}}
The previous section showed that the ray-tracing equations can describe the propagation of light from the emission regions to observers. However, up to this point, there has been no mention of imaging. Theoretically, black hole images are generated through the application of geometric optics, where the direction of light is utilized to determine the relative position on the image plane.

In this section, we establish the celestial sphere for observers using astrometric observables. It was first proposed for investigating the black hole shadow in a coordinate-independent and tetrad-independent manner \cite{Chang:2020miq}. Here, we will present an extension of the approach.

\

\subsection{Integral constants for finite distant observers}

To ensure that the light ray can reach the location of the observers $(r_{\rm o}, \theta_{\rm o})$, the following conditions must be satisfied,
\begin{eqnarray}
  \mathcal{R} (r_{\rm o})  \geqslant  0~, \hspace{0.5cm}  \Theta (\theta_{\rm o}) \geqslant  0 ~,
\end{eqnarray}
It can restrict the ranges of the constants $\kappa$ and $\lambda$ through the following inequality,
\begin{eqnarray}
  \Delta_{\theta} (\theta_{\rm o}) \kappa - \frac{(\lambda - a   \sin^2
    \theta_{\rm o})^2}{\sin^2 \theta_{\rm o}} & \geqslant & 0  \label{rangekappa}~,\\
  (r_{\rm o}^2 + a^2 - a \lambda)^2 - \Delta_r (r_{\rm o}) \kappa & \geqslant & 0~.  \label{rangelambda}
\end{eqnarray}
For observers approaching the outer horizon, the $\Delta_r
  (r_{\rm o})$ has an opposite sign separated by the outer horizon. 

For simplification, we introduce the polar coordinates $(\rho, \varphi)$, which are defined via
\begin{eqnarray}
  \rho^2 & \equiv & \Delta_{\theta} \kappa - (\lambda - a)^2 + \lambda^2 + a^2
  \cos^2 \theta_{\rm o}~, \\
  \cos \varphi & \equiv & - \frac{\lambda}{\rho \sin \theta_{\rm o}} ~,
\end{eqnarray}
such that Eq.~(\ref{rangekappa}) can reduce to an trival inequality
$\rho^2 \sin^2 \varphi \geqslant 0$, and Eq.~(\ref{rangelambda}) is evaluated to be
\begin{eqnarray}
  0&\geqslant&(a^2 \Delta_{\theta} \cos^2 \varphi \sin^2 \theta_{\rm o} - \Delta_r) \rho^2 - 2 a
  \cos \varphi \sin \theta_{\rm o} (\Delta_r - \Delta_{\theta} (r_{\rm o}^2 + a^2)) \rho \nonumber\\ && +
  (r_{\rm o}^2 + a^2)^2 \Delta_{\theta} - a^2 \Delta_r \sin^2 \theta_{\rm o}  ~. \label{ineq2}
\end{eqnarray}
For distant observers situated beyond the outer horion $r_{\rm o} \rightarrow
  \infty$, the above equation reduces to
\begin{eqnarray}
  (\rho + {\rm acos} \varphi \cos \theta)^2 + \frac{3}{\Lambda} + a^2 (1 -
  \cos^2 \varphi \sin^2 \theta) & \geqslant & 0 ~.
\end{eqnarray}
It is evident that the inequality in Eq.~(\ref{ineq2}) holds as $r_{\rm o} \rightarrow\infty$, implying no constraint on $\rho$ and $\varphi$ in this case. However, it does not indicate that the images of a Kerr-de Sitter black hole are well-defined for all distant observers. The counterexample is that the 4-velocities of static distant observers beyond the outer horizon can not be obtained \cite{Perlick:2018iye, Chang:2019vni}. Thus, the above result shows that the ranges of $\rho$ and $\varphi$ can only determine the reach of light, but they are not direct observables.

\

\subsection{Astrometric approaches}

Since $\rho$ and $\varphi$ are mathematically derived from the integral constants $\kappa$ and $\lambda$, the images represented by $(\rho, \varphi)$ are not direct observables in general. Currently, two distinct approaches can be employed to elucidate the black hole images that we observe. The first approach involves establishing a local reference frame at an observer's location, where a set of ideal tetrads is used to locate the light rays. For simplicity, the local frame of zero-angular-momentum observers is often adopted for studying shadows \cite{Bardeen:1973tla} or for conducting ray-tracing \cite{Cardenas-Avendano:2022csp}. However, for finite distant obverser, this frame does not move geodesically, making it somewhat artificial. The second approach involves locating light rays using astrometric observables \cite{Chang:2020miq}. Previous studies have shown that the shadow can be sketched with the relative angular distance to a set of reference light rays. Thus, it is in a tetrad-independent manner. However, this approach has only been applied for calculating the shadow to date \cite{Chang:2020miq, Chang:2021ngy, Chang:2020lmg, He:2020dfo}.

As mentioned above, the astrometric approach relies on the choices of
reference light rays. In the previous studies, the authors selected the light
rays with its 4-momentums, $p_{\theta}|_{\bm x_{\rm o}} = 0$. Namely, the references are two light rays originating from the edges of the photon sphere, which can be formulated as
\begin{eqnarray}
  k & = & p|_{(\kappa (r_{{\rm c}, \min}), \lambda (r_{{\rm c}, \min}))} ~,\\
  w & = & p|_{(\kappa (r_{{\rm c}, \max}), \lambda (r_{{\rm c}, \max}))} ~,
\end{eqnarray}
where the 4-momentum of light $p_{\mu}$ has been given in Eqs.~(\ref{mom:p}), and $r_{\rm c}$ denotes the locaion of photon sphere. With the criterions for critical curves, namely, $\mathcal{R} (r) =\mathcal{R}' (r) = 0$,
the integral constants $\kappa$ and $\lambda$ for the light rays on the photon sphere can be given by
\begin{eqnarray}
  \kappa (r_{\rm c}) & = & - \frac{12 r_{\rm c}^2 (a^2 (r_{\rm c}^2 \Lambda - 3) + r_{\rm c} (r_{\rm c}^3
    \Lambda - 3 r_{\rm c} + 6 M))}{(3 M + r_{\rm c} (2 r_{\rm c}^2 \Lambda + a^2 \Lambda - 3))^2}
  ~,\\
  \lambda (r_{\rm c}) & = & \frac{3 r_{\rm c}^2 (r_{\rm c} - 3 M) + a^4 r_{\rm c} \Lambda + a^2 (r^3_{\rm c}
  \Lambda + 3 r_{\rm c} + 3 M)}{a (3 M + r_{\rm c} (2 r_{\rm c}^2 \Lambda + a^2 \Lambda - 3))} ~.
\end{eqnarray}
And the minimum and maximum of $r_{\rm c}$ can be determined by $\Theta (\theta_{\rm o})
  |_{(\kappa (r_{\rm c}), \lambda (r_{\rm c}))} = 0$ for given
$\theta_{\rm o}$. Because the equation of $\Theta (\theta_{\rm o})$ is a sextic equation with respect to the $r_{\rm c}$, here, the $r_{{\rm c}, \min}$ and $r_{{\rm c}, \min}$ are obtained numerically.

\begin{figure}
  \includegraphics[width=.6\linewidth]{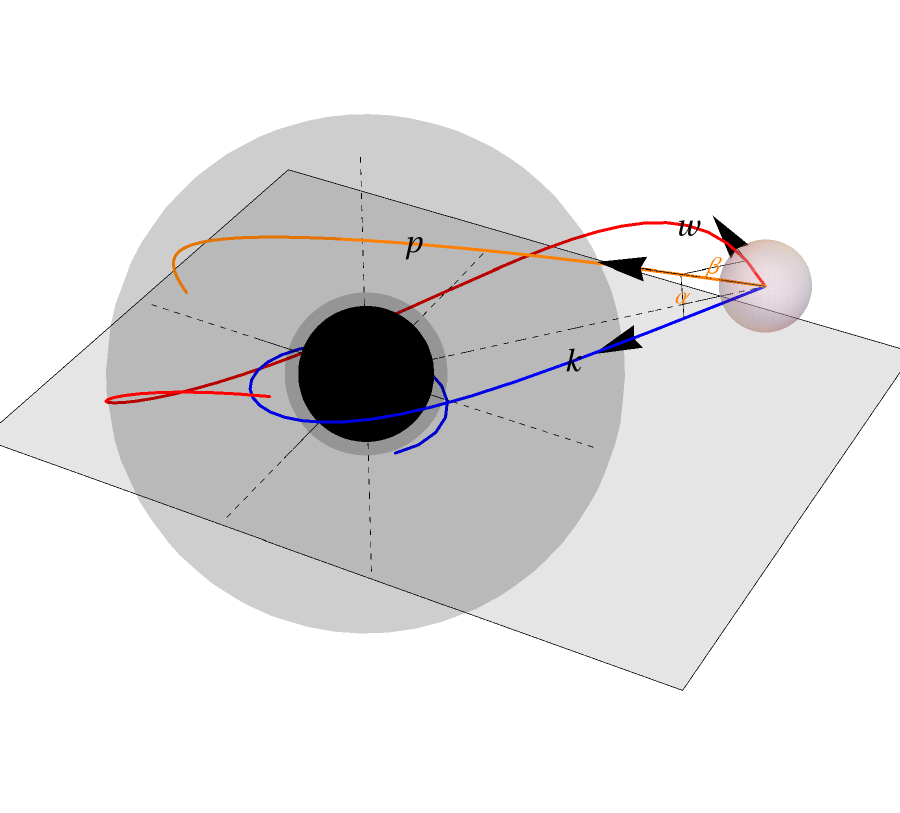}  
  \caption{A schematic diagram illustrating the astrometric approach for locating light ray $p$ with respect to the reference $w$ and $k$. \label{F0}}
\end{figure}

The angle between two light rays $a$ and $b$ is also known as astrometric observable \cite{Chang:2020miq,alma991000929205807309}, and can be defined with
\begin{eqnarray}
  \psi & = & {\rm Angle} (a, b) \equiv \arccos \left(
  \frac{\gamma^{*} a \cdot \gamma^{*} b}{| \gamma^{*}
    a |   | \gamma^{*} a |} \right) = {\rm arcos} \left( 1 +
  \frac{a \cdot b}{(u \cdot a) (u \cdot b)} \right) ~,
\end{eqnarray}
where $\gamma^{*}$ is spatial induced metric, $u$ is the 4-velocities
of observers, and the $\cdot$ represent the index contraction with metric
in Eq.~(\ref{met}). Thus, it is in a coordinate-independent manner. With the reference light rays $k$ and $w$, we can locate the relative position of light ray $p$ via
\begin{eqnarray}
  \alpha & = & {\rm Angle} (p, k) ~,\\
  \beta & = & {\rm Angle} (p, w) ~,
\end{eqnarray}
The schematic diagram is presented in Figure~\ref{F0}.
Utilizing the angular distance $\alpha, \beta$ and $\gamma(\equiv{\rm Angle} (k, w))$, the location of $p$ can be rewritten as celestial coordinates,
\begin{eqnarray}
  \Psi & = & \arccos \left( \sin \beta \sqrt{1 - \left( \frac{\cos \alpha -
      \cos \beta \cos \gamma}{\sin \beta \sin \gamma} \right)^2} \right) ~,\\
  \Phi & = &  \arccos \left( \frac{\cos \beta}{\sin \Psi} \right) ~.
\end{eqnarray}
However, the above formalism can not distinguish between the northern and southern hemispheres on the celestial sphere. In previous studies \cite{Chang:2020miq, Chang:2021ngy}, the authors employed two distinct parameter regions, $\varphi \in [0, \pi]$ and $\varphi \in [\pi, 2\pi]$, to distinguish between the hemispheres. However, this trick is only applicable for sketching the shadow. Therefore, here, we introduce the third reference light rays $l$
on the northern hemispheres, and obtain the criterion for determining which hemisphere of a light ray is located, namely,
\begin{eqnarray}
  \Upsilon & = & {\rm sign} (\cos \delta - \cos (\Phi - \Phi_l) \sin \Psi
  \sin \Psi_l)~,
\end{eqnarray}
where $\delta = {\rm Angle} (p, l)$, $\Phi_l \equiv \Psi |_{p = l}$ and $\Psi_l \equiv
  \Psi |_{p = l}$. If a light ray is located in the northern hemisphere, we have $\Upsilon=1$. 
  
Additionally, to show the results in a 2-dimensional plane, we utilize the stereographic projection,
\begin{eqnarray}
  Y & = & -\frac{2 \sin \Phi \sin \Psi}{1 + \cos \Phi \sin \Psi}~,\\
  Z & = & \frac{2 \cos \Psi}{1 + \cos \Phi \sin \Psi}~.
\end{eqnarray}
The images on the projection plane $(Y,Z)$ might have unphysical image distortion if the observers are too close to the black hole. In this case, the reliable results should be given with the celestial coordinate $(\Psi,\Phi)$.

Besides the new equalities $\delta$, $\Upsilon$ and reference light $l$, the framework of the astrometric approach has already been introduced in previous studies \cite{Chang:2020miq, Chang:2021ngy}.

\

\section{Lower-order images for finite-distant observers \label{V}}

Due to the aberration of light, the shadow can be distorted or resized for finite-distant observers in motion \cite{Chang:2021ngy}. In this study, we extend the investigation into the behavior of primary, secondary, and $n=2$ images, associating it with Penrose's aberration formula \cite{1959PCPS...55..137P}. Previous studies have shown that co-moving observers at spatial infinity can still observe the shadow \cite{Perlick:2018iye, Chang:2019vni, Li:2020drn}. Moreover, the shadow of a Kerr black hole appears round in the view of co-rotating observers \cite{Chang:2020lmg}. Therefore, we anticipate that non-trivial outcomes for the lower-order images can also be found.

To investigate the aberration effect, our starting point is to consider observers in motion, which is generally described by the 4-velocities of timelike geodesic, namely,
\begin{subequations}
  \begin{eqnarray}
    u^t & = & \mathcal{E} \left( \frac{1}{N} - \frac{A}{G_3} (A + \mathcal{L})
    \right)~,\\
    u^r & = & \frac{\sigma_r}{\Sigma} \sqrt{\mathcal{E}^2 (f_1 (r ; \mathcal{L}) -
    \mathcal{K}) - r^2}~,\\
    u^{\theta} & = & \frac{\sigma_{\theta}}{\Sigma} \sqrt{\mathcal{E}^2 (f_2
    (\theta ; \mathcal{L}) + \mathcal{K}) - a^2 \cos^2 \theta}~,\\
    u^{\phi} & = & \frac{\mathcal{E} (A + \mathcal{L})}{G_3}~,
  \end{eqnarray}\label{u:gen}
\end{subequations}
where $\sigma_r, \sigma_\theta = \pm$, the above quantities are defined with $N  \equiv  - g_{00}$, $G_3  \equiv  g_{33} - {g_{03}^2}/{g_{00}}$,  $A \equiv  {g_{03}}/{g_{00}} $, and 
\begin{subequations}
  \begin{eqnarray}
    f_1(r;\mathcal{L}) & = & \frac{(a^2 + r^2 - a \mathcal{L})^2}{\Delta_r}~,\\
    f_2(\theta;\mathcal{L}) & = & - \frac{1}{\Delta_{\theta}} \left( a {\rm sin} \theta -     \frac{\mathcal{L}}{\sin \theta} \right)^2~.
  \end{eqnarray}
\end{subequations}
Here, we adopt the parameterization scenario for the timelike 4-velocities used in Ref.~\cite{Chang:2021ngy}. There are three integral constants $\mathcal{K}$, $\mathcal{L}$, and $\mathcal{E}$.
As Penrose's aberration formula was to quantify the apparent shapes and sizes of a rigid sphere, we also strict our attention on the shapes and sizes of the primary, secondary and $n=2$ images of the thin accretion disk. For illustration, we consider $r_{\rm d,-}=4M$ and $r_{\rm d,+}=10M$ in Eq.~(\ref{Iemt}) as the representative case.

\

\subsection{The images in the view of static observers and co-moving observers \label{V.A}}

\begin{figure}
  \includegraphics[width=0.8\linewidth]{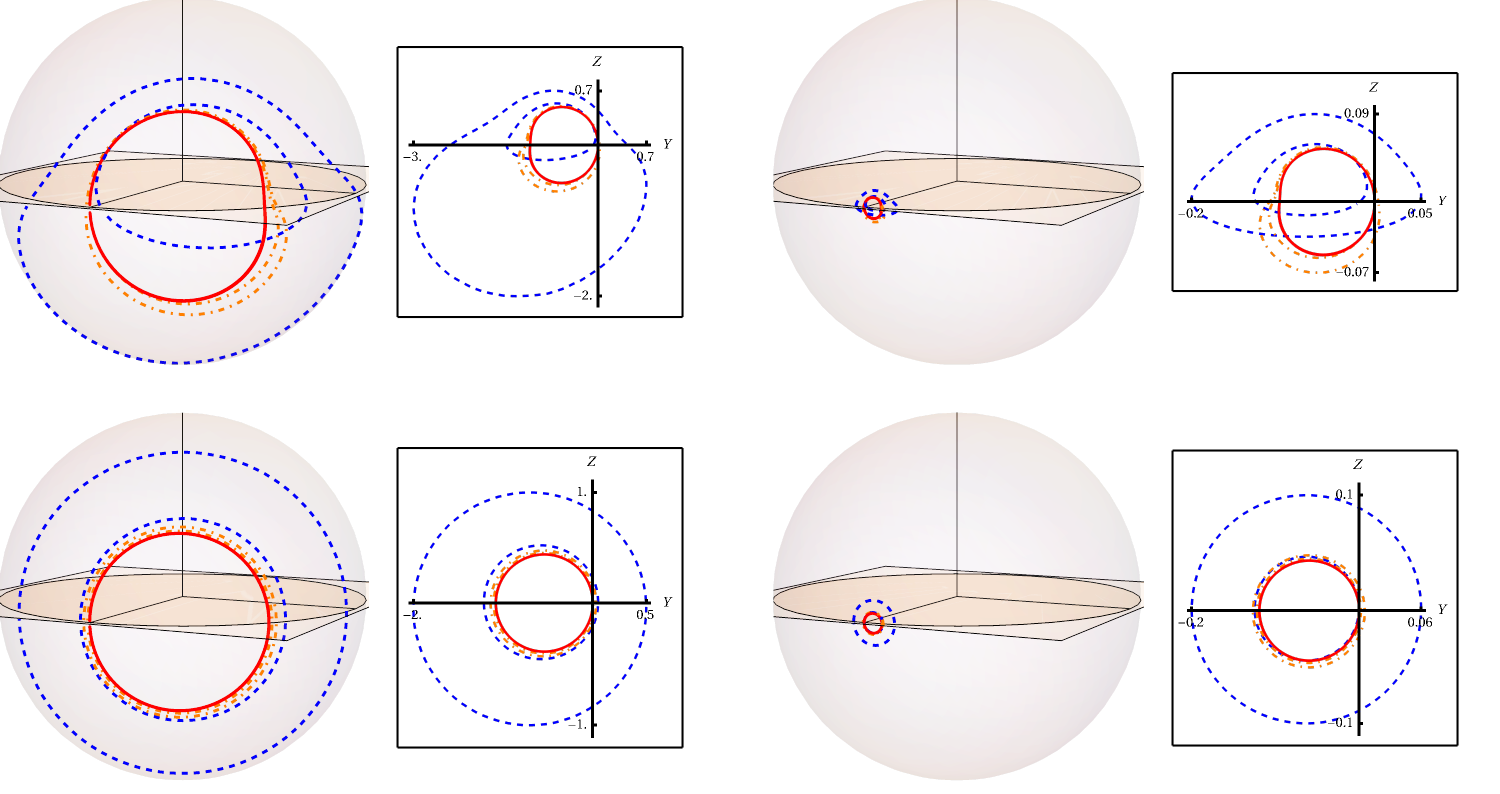}
  \caption{
    Primary (dashed), secondary (dashed-dotted), and $n=2$ images of the thin accretion disk around a Kerr black hole with a spin parameter $a=0.99M$ for selected static observers on the celestial sphere and projection plane. The observers are located at $r_{\rm o}=10M$, $\theta_{\rm o}=2\pi/5$ (top-left panel), $r_{\rm o}=10M$, $\theta_{\rm o}=\pi/5$ (top-right panel), $r_{\rm o}=100M$, $\theta_{\rm o}=\pi/25$ (bottom-left panel), and $r_{\rm o}=100M$, $\theta_{\rm o}=\pi/25$ (bottom-right panel), respectively.\label{F1}}
\end{figure}

\begin{figure}
  \includegraphics[width=.8\linewidth]{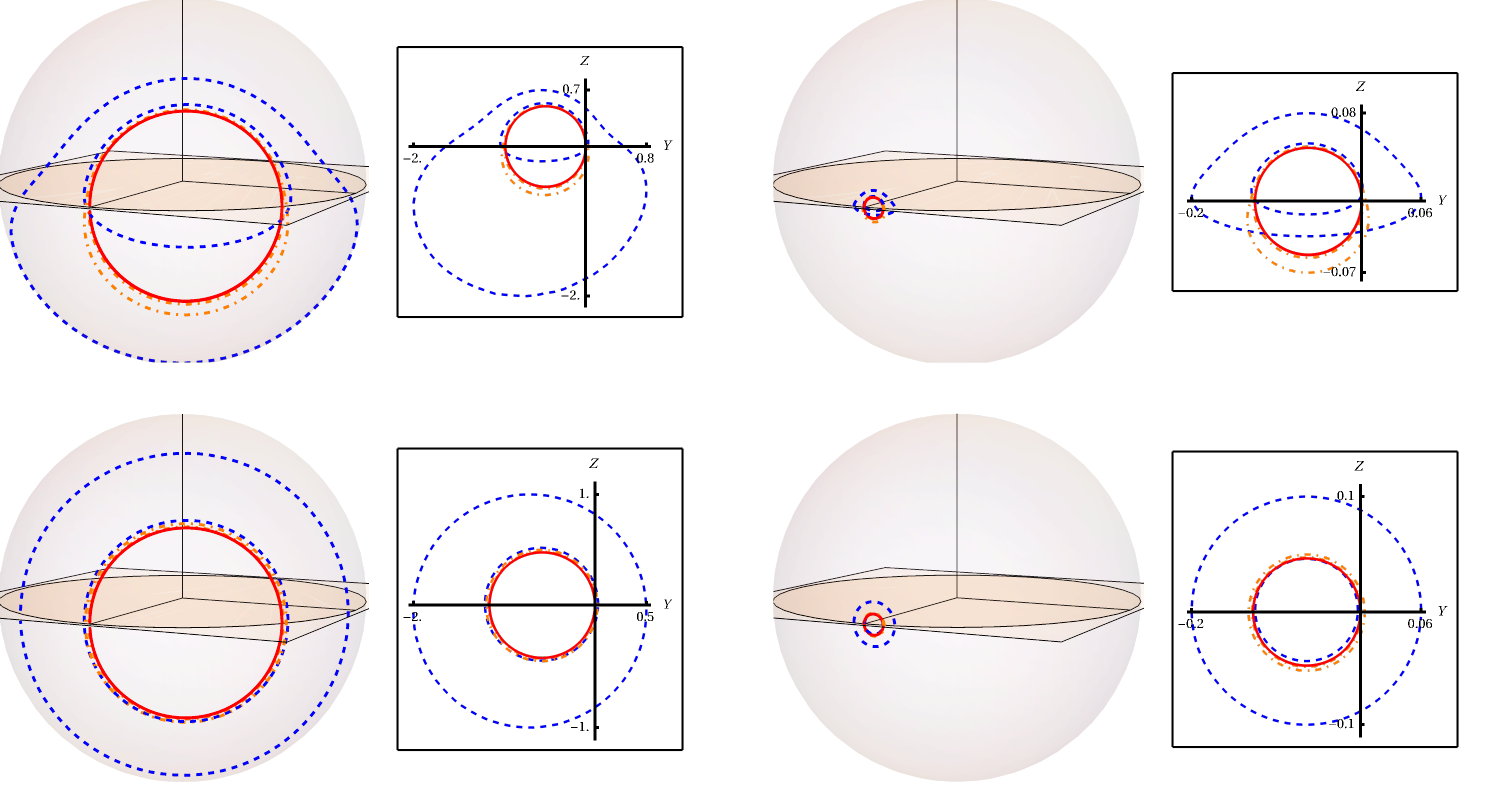}
  \caption{Primary (dashed), secondary (dashed-dotted) and $n$=2 images of the thin accretion disk around a Kerr black hole with spin parameter $a=0.1M$ for selected static observers on the celestial sphere and projection plane. The observers are located at $r_{\rm o}=10M$, $\theta_{\rm o}=2\pi/5$ (top-left panel), $r_{\rm o}=10M$, $\theta_{\rm o}=2\pi/5$ (top-right panel), $r_{\rm o}=100M$, $\theta_{\rm o}=\pi/25$ (bottom-left panel), and  $r_{\rm o}=100M$, $\theta_{\rm o}=\pi/25$ (bottom-right panel), respectively. \label{F2}}
\end{figure}

To present the variation of the lower-order images for observers in motions, we sketch the images for static observers and co-moving observers at first. The 4-velocities of static observers can be given by
\begin{eqnarray}
  u_{{\rm stc}} & = & \left( \frac{1}{\sqrt{N}}, 0, 0, 0 \right)~.
\end{eqnarray}
The above 4-velocity is not a tangent vector of geodesics. It should be understood as instantaneous velocities with respect to a specific location.
Figure~\ref{F1} and \ref{F2} show the lower-order images in the view of near and distant static observers for given black hole parameters. For observers at inclination angle $\theta_{\rm o}=2\pi/5$, the primary images in the view of near observers are enlarged, and cover the secondary and $n=2$ images.

\begin{figure}
  \includegraphics[width=.8\linewidth]{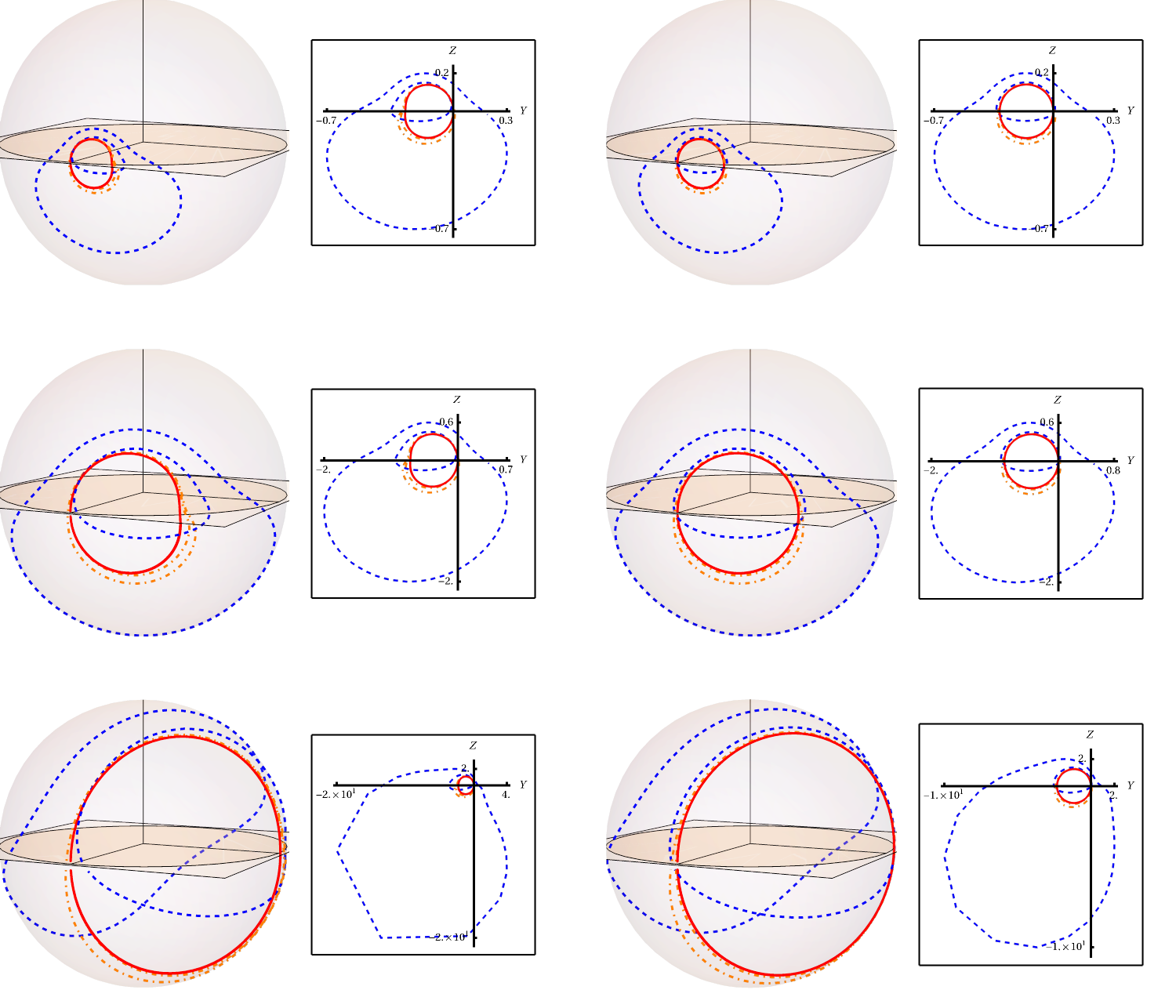}
  \caption{Primary (dashed), secondary (dashed-dotted), and $n$=2 images of the thin accretion disk around a Kerr-de Sitter black hole with a cosmological constant $\Lambda=0.01M^{-2}$ and spin parameters $a=0.99M$ (left panels) and $a=0.1M$ (right panels). The observers are set as in-going freely-falling observers (top panels), static observers (center panels), and out-going co-moving observers (bottom panels), all inclined at an angle $\theta_{\rm o}=2\pi/5$.  \label{F3}}
\end{figure}

\begin{figure}
  \includegraphics[width=.8\linewidth]{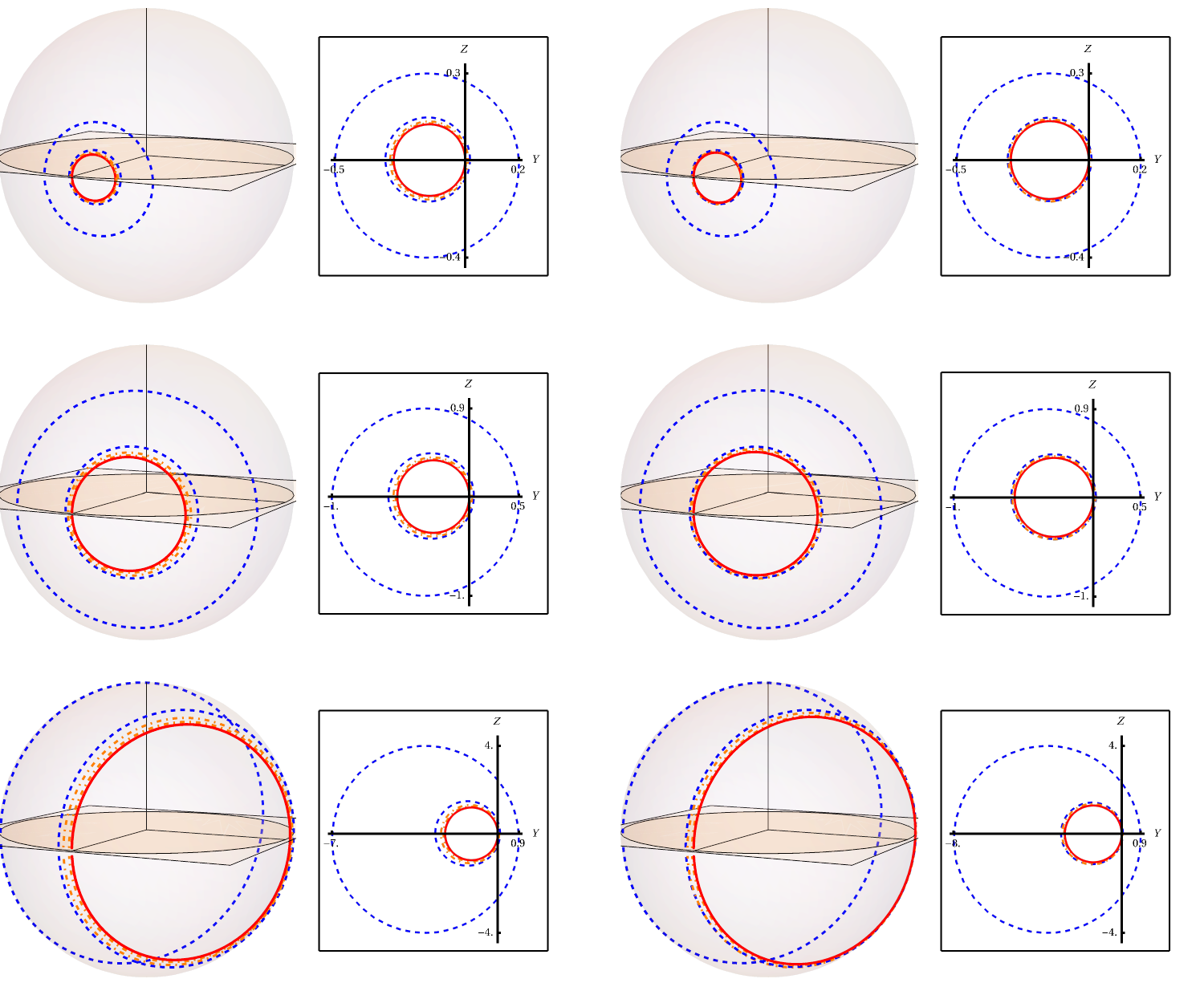}
  \caption{Primary (dashed), secondary (dashed-dotted), and $n$=2 images of the thin accretion disk around a Kerr-de Sitter black hole with a cosmological constant $\Lambda=0.01M^{-2}$ and spin parameters $a=0.99M$ (left panels) and $a=0.1M$ (right panels). The observers are in-going freely-falling observers (top panels), static observers (center panels), and out-going co-moving observers (bottom panels), all positioned at the inclination angle $\theta_{\rm o}=\pi/25$. \label{F4}}
\end{figure}

In the presence of the cosmological constant, consideration of images with respect to co-moving observers is necessary. This is because i) there is no doubt that the co-moving frame serves as the physical reference frame, and ii) the 4-velocity $u_{{\rm stc}}$ is no longer well-defined beyond the outer horizon. The latter implies that black hole images can not be obtained for static observers.
Letting the integral constants to be $\kappa = a^2,$ $\lambda = 0$ and
$\mathcal{E}= 1$, we have
\begin{subequations}
  \begin{eqnarray}
      u_c & = & \left( \frac{1}{N} - \frac{A^2}{G_3}, \sigma_r \frac{\sqrt{(f_1 (r ; 0) -
      (a^2 + r^2)) \Delta_r}}{\Sigma}, \frac{\sqrt{(f_2 (r ; 0) + a^2 \sin^2
      \theta) \Delta_{\theta}}}{\Sigma}, \frac{A}{G_3} \right)~. \label{u:comov}
  \end{eqnarray}
\end{subequations}
If letting $\sigma_r = - 1$, we have an in-going 4-velocity. It can describe the motion of freely-falling observers. The freely falling observers are shown to be frozen at both the inner and outer horizons. One can check the coordinate
3-velocities $u^r / u^t = u^{\theta} / u^t = 0$ at the horizons. Despite this, it does not indicate that the freely-falling observers are not well-defined when crossing the horizon. One can obtain correct trajectories using Mino time, for
examples. In the case of $\Lambda = a = 0$, the co-moving observer moves
radially. And for distant observers $r\rightarrow\infty$, we have
\begin{subequations}
  \begin{eqnarray}
    u^r_c & = & r \sqrt{\frac{\Lambda}{3}} +\mathcal{O} (1)~,\\
    u^{\theta}_c = u^{\phi}_c & = & \mathcal{O} \left( \frac{M^2}{r^2} \right)~.
  \end{eqnarray}
\end{subequations}
From the 4-velocity, one can read Hubble's law in the low-reshift limit, since redshift $z
  \simeq u^r$ and Hubble's constant $H_0 \equiv \sqrt{\Lambda / 3}$ in de-Sitter
space-time.
Figure~\ref{F3} and \ref{F4} show the images of Kerr-de Sitter black hole for in(out)-going freely-falling and static observers. The out-going observers would see a larger image of the emission, which is consistent with the picture that co-moving observers in the spatial infinity could still see the shadow \cite{Perlick:2018iye,Chang:2019vni}.  In the bottom panels of \ref{F3} and \ref{F4}, one might find the unphysical image distortion on the projection plane $(Y,Z)$. In this case, reliable results of the images should be given with the celestial coordinate $(\Psi,\Phi)$.

\

\subsection{Observing Black hole images with moving-telescope networks \label{V.B}}

The consideration of static observers and co-moving observers does not encompass all observers in motion. These particular observers were chosen in the preceding sections because they serve as references for defining relative motions.

Suppose telescopes are launched from the reference frame with respect to the static observers or co-moving observers, and they are launched with high speed. Each telescope might observe different images of the emissions. The key question is whether the distinctions in the images induced by the aberration effect are simply kinematic effects or can reflect the spacetime geometries.

Here, we introduce the relative 3-speed of $u$ in the form of
\begin{equation}
  \upsilon  \equiv  \frac{| \gamma^{*} u |}{u_{\rm ref} \cdot u} = \sqrt{1 - \frac{1}{(u_{\rm ref} \cdot u)^2}}~,
\end{equation}
where $u_{\rm ref}$ is the 4-velocity of the reference frame.
We would clarify that the quantity $\upsilon$ here is employed for distinguishing 4-velocities $u$ in a coordinate-independent manner. And we do not assume that it has robust physical meanings so far.

\begin{figure}
  \includegraphics[width=.95\linewidth]{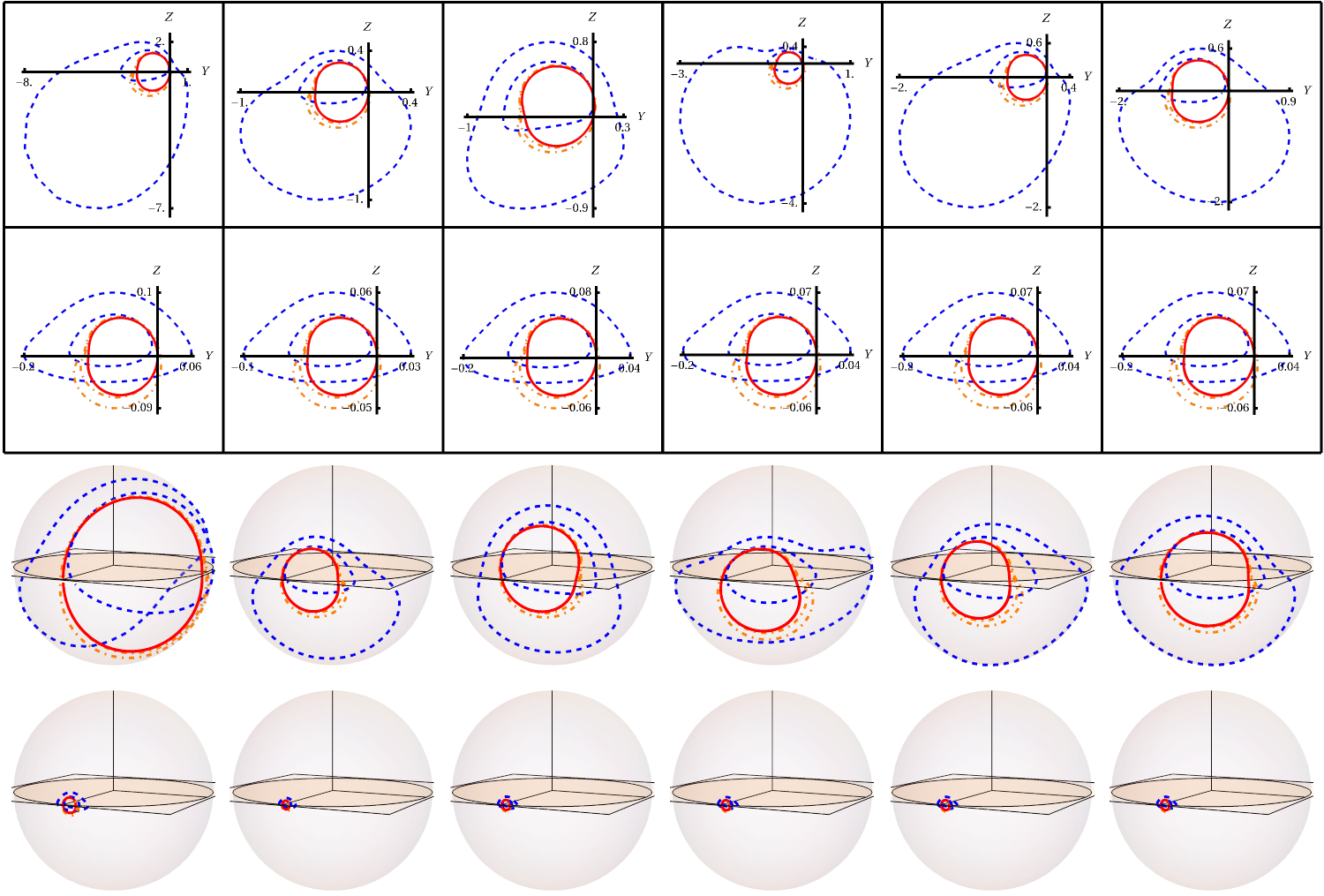}
  \caption{Primary (dashed), secondary (dashed-dotted), and $n$=2 images of the thin accretion disk around a Kerr black hole with a spin parameter $a=0.99M$ for selected moving observers with respect to the static frame. Specifically, the moving observers are located at distances $r_{\rm o}=10M$ and $100M$, with the inclination angle $\theta_{\rm o}=2\pi/5$ and 4-velocities $u^{(r, \pm, \rm s)}$, $u^{(\theta, \pm, \rm s)}$, and $u^{(\phi, \pm, \rm s)}$ at a speed of $\upsilon=0.5$, respectively.  \label{F5}}
\end{figure}

\begin{figure}
  \includegraphics[width=.95\linewidth]{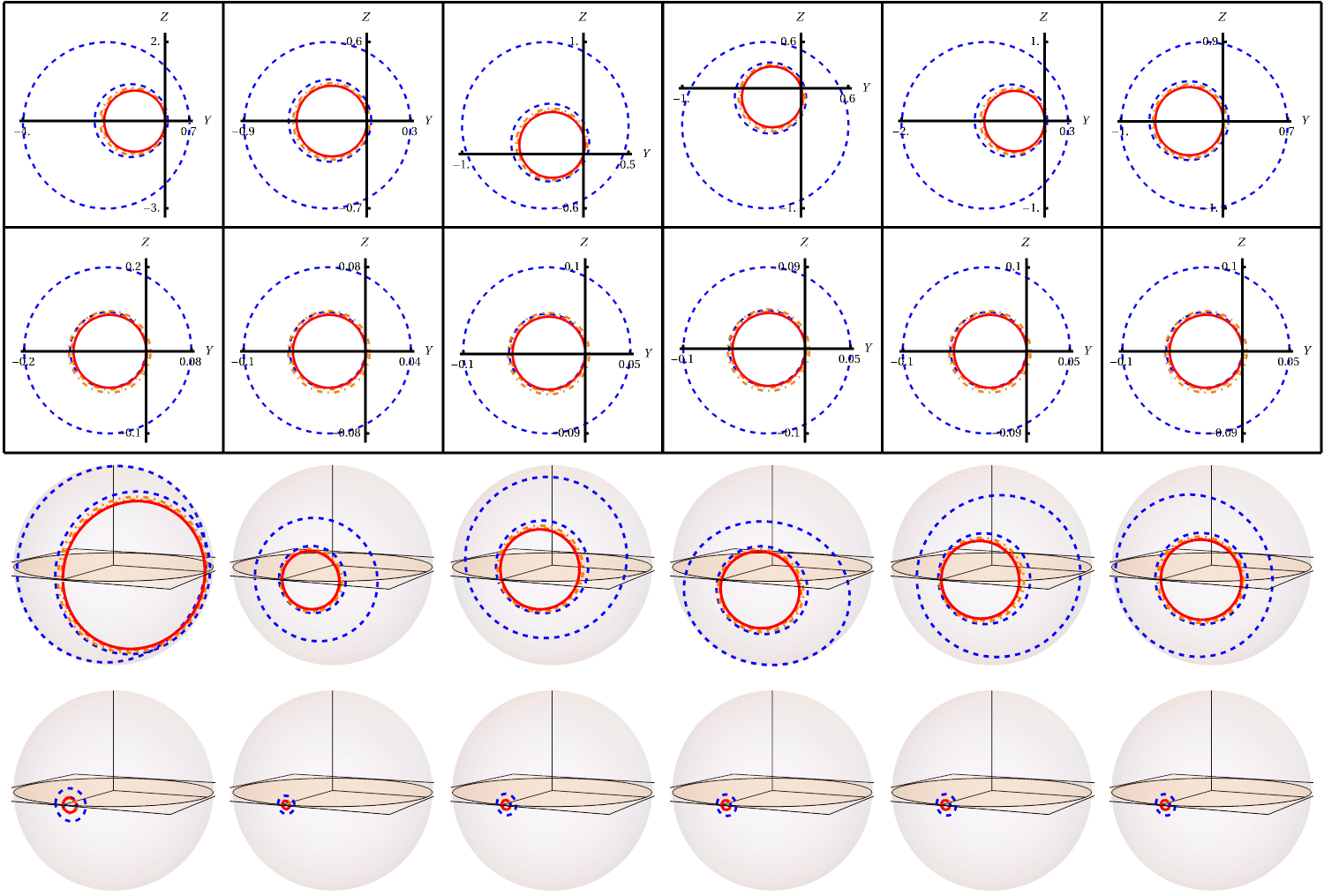}
  \caption{Primary (dashed), secondary (dashed-dotted), and $n$=2 images of the thin accretion disk around a Kerr black hole with a spin parameter $a=0.99M$ for selected moving observers with respect to the static frame. Specifically, the moving observers are located at distances $r_{\rm o}=10M$ and $100M$, with the inclination angle $\theta_{\rm o}=\pi/25$ and 4-velocities $u^{(r, \pm, \rm s)}$, $u^{(\theta, \pm, \rm s)}$, and $u^{(\phi, \pm, \rm s)}$ at a speed of $\upsilon=0.5$, respectively. \label{F6}}
\end{figure}

Figures~\ref{F5} and \ref{F6} show the images in the view of near and distant observers in motion with respect to the static frame.
The 4-velocities of moving observers are defined with
\begin{subequations}
  \begin{eqnarray}
    u^{(r, \pm, \rm s)} & = & \mathcal{E} \left( \frac{1}{N}, \pm
    \sqrt{\frac{\Delta_r}{\Sigma} \left( \frac{1}{N} - \frac{1}{\mathcal{E}^2}
    \right)}, 0, 0 \right)~,\\
    u^{(\theta, \pm, \rm s)} & = & \mathcal{E} \left( \frac{1}{N}, 0, \pm
    \sqrt{\frac{\Delta_{\theta}}{\Sigma} \left( \frac{1}{N} -
    \frac{1}{\mathcal{E}^2} \right)}, 0 \right)~,\\
    u^{(\phi, \pm, \rm s)} & = & \left( \frac{\mathcal{E}}{N} \mp \frac{A}{G_3}
    \sqrt{G_3 \left( \frac{\mathcal{E}^2}{N} - 1 \right)}, 0, 0, \pm
    \frac{1}{G_3} \sqrt{G_3 \left( \frac{\mathcal{E}^2}{N} - 1 \right)} \right)~.\end{eqnarray}\label{u:int:stc}
\end{subequations}
The 4-velocities are also instantaneous velocities and are derived from a specific choice of $\kappa$ and $\lambda$, depending on a given $\bm x_{\rm o}$. For near observers, the shapes of the primary images become highly distorted compared to the secondary and $n=2$ images. Additionally, a tailing behavior is observed along the direction of motion. This behavior can be further illustrated in the case of observers at the inclination angle $\theta_{\rm o}=\pi/25$. Specifically, the outer edge of primary images is shown to be less sensitive to the speed $\upsilon$ compared to the inner one. Furthermore, for observers at the same speed $\upsilon$ located far away from the black holes, the distortion of primary images is shown to be suppressed. It might indicate that the aberration effect is influenced by the gravity environment of the observers. Previous studies on black hole shadow also obtained similar results \cite{Chang:2021ngy}.

Similarly, we consider the 4-velocities of moving observers with respect to the co-moving frame, which are defined with
\begin{subequations}
{\small  \begin{eqnarray}
    u^{(r, \pm, \rm c)} & = & \left( \frac{\mathcal{E}}{N} - \frac{A^2}{G_3}, \pm
    \frac{\sqrt{\Delta_r (f_2 (\theta ; 0) - (a^2 + r^2) + N^{- 1} \Sigma
    (\mathcal{E}^2 - 1))}}{\Sigma}, \frac{\sqrt{\Delta_{\theta} (f_2 (\theta ;
    0) + a^2 \sin^2 \theta)}}{\Sigma}, 0 \right)~,\nonumber\\ &&\\
    u^{(\theta, \pm, \rm c)} & = & \left( \frac{\mathcal{E}}{N} - \frac{A^2}{G_3},
    \frac{\sqrt{\Delta_r (f_2 (\theta ; 0) - (a^2 + r^2))}}{\Sigma}, \pm
    \frac{\sqrt{\Delta_{\theta} (f_2 (\theta ; 0) + a^2 \sin^2 \theta + N^{- 1}
    \Sigma (\mathcal{E}^2 - 1))}}{\Sigma}, 0 \right)~,\nonumber\\ &&\\
    u^{(\phi, \pm, \rm c)} & = & \left( \frac{\mathcal{E}}{N} \mp\frac{A \sqrt{N (G_3
    (\mathcal{E}^2 - 1) + A^2 N)}}{G_3 N}, \frac{\sqrt{\Delta_r (f_2 (\theta ;
    0) - (a^2 + r^2))}}{\Sigma}, \frac{\sqrt{\Delta_{\theta} (f_2 (\theta ; 0) +
    a^2 \sin^2 \theta)}}{\Sigma}, \right. \nonumber \\ && \left. \pm \frac{\sqrt{N (G_3 (\mathcal{E}^2 - 1) + A^2
    N)}}{G_3 N} \right)~.
  \end{eqnarray}} \label{u:int:com}
\end{subequations}
The above instantaneous velocities can provide a picture of the establishment of moving telescope networks. It is known that our Earth is co-moving with respect to distant supermassive black holes, such as the M87. By launching telescopes into space with different velocities formulated by Eqs.~(\ref{u:int:com}), one can test whether the distinctions in the images from different telescopes are consistent with the theoretical predictions here, in principle.

\begin{figure}
  \includegraphics[width=1\linewidth]{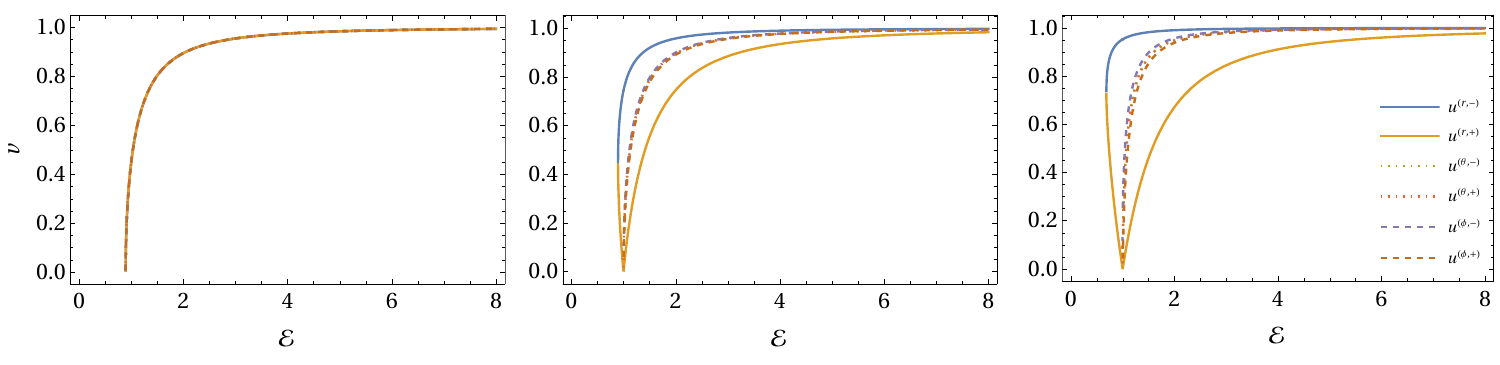}
  \caption{Left panel: The $\mathcal{E}$-$\upsilon$ relations for 4-velocity with respect to static frame near a Kerr black hole with $a=0.99M$. Center panel: The $\mathcal{E}$-$\upsilon$ relations for 4-velocity with respect to the co-moving frame near a Kerr black hole with $a=0.99M$. Right panel: The $\mathcal{E}$-$\upsilon$ relations for 4-velocity with respect to the co-moving frame near a Kerr-de Sitter black hole with $a=0.99M$ and $\Lambda=0.01M^{-2}$. All these frame locates at $r_{\rm o}=10M$ and $\theta_{\rm o}=2\pi/5$. \label{F7}}
\end{figure}

\begin{figure}
  \includegraphics[width=1\linewidth]{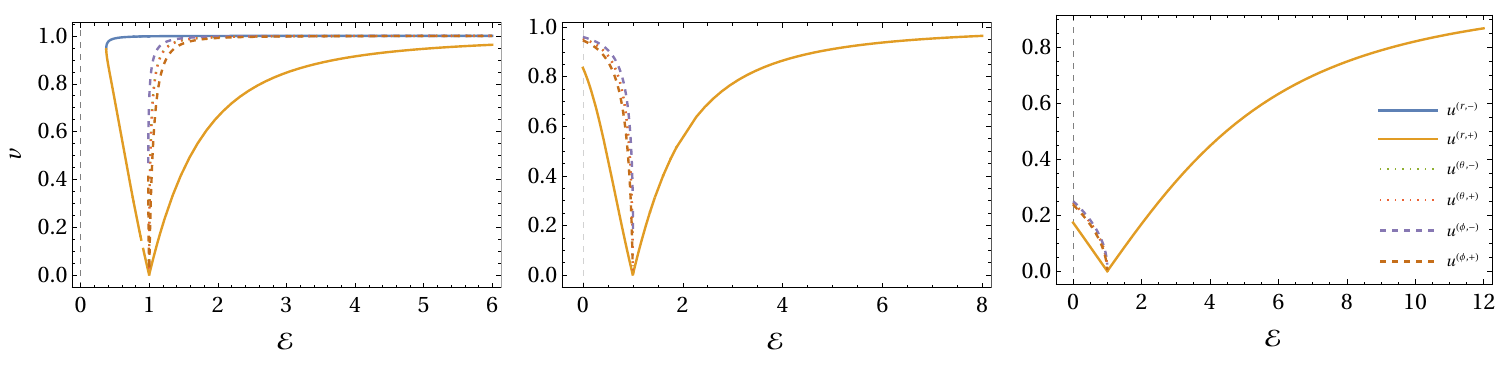}
  \caption{The $\mathcal{E}$-$\upsilon$ relations for observers located at $r_{\rm o}=15M$ (Left panel), $20M$, (Center panel), and $100M$ (Right panel). In all these cases, co-moving frames are situated at $\theta_{\rm o}=2\pi/5$ in a Kerr-de Sitter space-time with the spin of $a=0.99M$ and the cosmological constant of $\Lambda=0.01M^{-2}$, resulting in an outer horizon at $r_{\rm H}\simeq16.2M$. \label{F8}}
\end{figure}

\begin{figure}
  \includegraphics[width=1\linewidth]{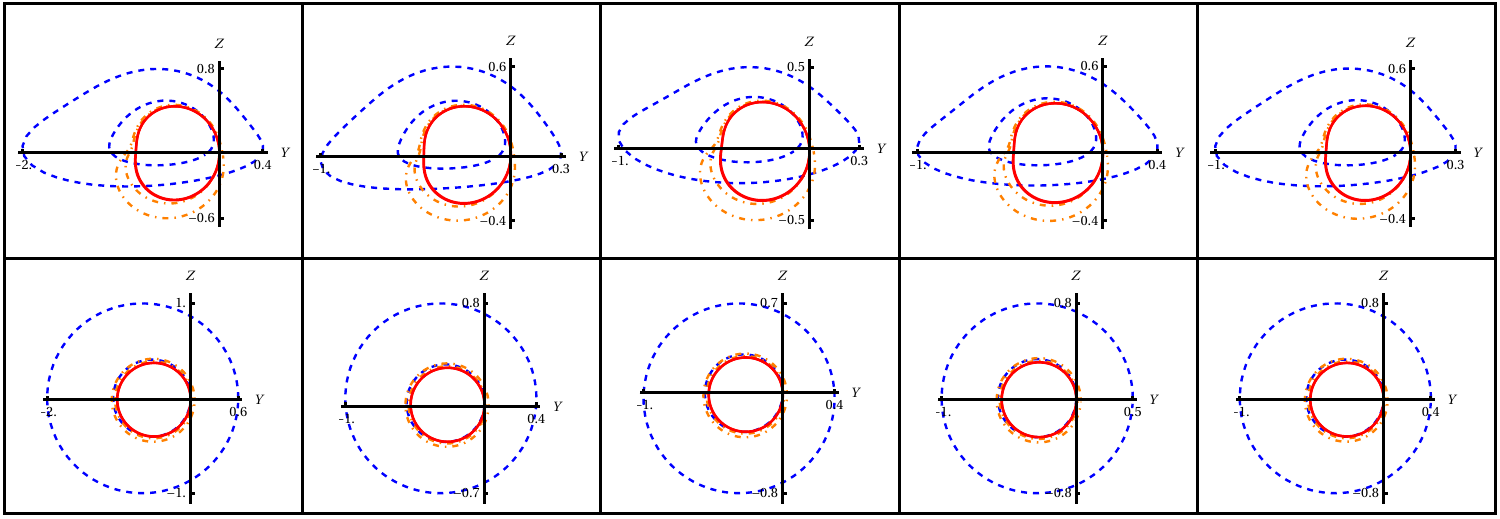}
  \caption{Primary (dashed), secondary (dashed-dotted), and $n$=2 images of the thin accretion disk around a Kerr-de Sitter black hole with the spin parameter of $a=0.99M$ and the cosmological constant of $\Lambda=0.01M^{-1}$ for moving observers. Specifically, the moving observers are located at the distance of $r_{\rm o}=100M$ and the inclination angle of $\theta_{\rm o}=2\pi/5$ (top panels) and $\theta_{\rm o}=\pi/25$ (bottom panels) with 4-velocities $u^{(r, +, \rm c)}$, $u^{(\theta, \pm, \rm c)}$, and $u^{(\phi, \pm, \rm c)}$ at a speed of $\upsilon=0.2$, respectively. \label{F9}}
\end{figure}

The initial launching speed $\upsilon$ is determined by a constant $\mathcal{E}$. To determine the proper $\mathcal{E}$, even beyond the outer horizon, we establish relations between the quantities $\upsilon$ and $\mathcal{E}$. Figure~\ref{F7} displays the $\mathcal{E}$-$\upsilon$ relations for different 4-velocities presented in Eqs.~(\ref{u:int:stc}) and (\ref{u:int:com}). It shows that for observers in the vicinity of the Kerr black hole or within the outer horizon of the Kerr-de Sitter black hole, it is always possible to launch telescopes at a 3-speed $\upsilon$ within and approaching the speed of light. It might enhance our confidence in regarding $\upsilon$ as a physical quantity.
Figure~\ref{F8} depicts the $\upsilon$-$\mathcal{E}$ relations for observers located at different distances from Kerr-de Sitter black holes. It is found that there are upper bounds on $\upsilon$ for the axial and in-going velocities if the observers are beyond the outer horizons. Further details on the upper bound will be discussed in the final part of this section.
In Figure \ref{F9}, we present the lower-order images for distant observers in motion with respect to the co-moving frame. Despite the observers being situated in a non-flat space-time, the aberration effect on the images is suppressed by the distance. Additionally, compared with the distant observers in Figure~\ref{F5}, it is evident that in the presence of a cosmological constant, i) the size of primary and secondary images is smaller, and ii) the distortion of the images is more sensitive to the motions of observers. The latter suggests that the observer-dependence of the images would be more significant than expected if the expansion of the universe is considered.

\begin{figure}
  \includegraphics[width=1\linewidth]{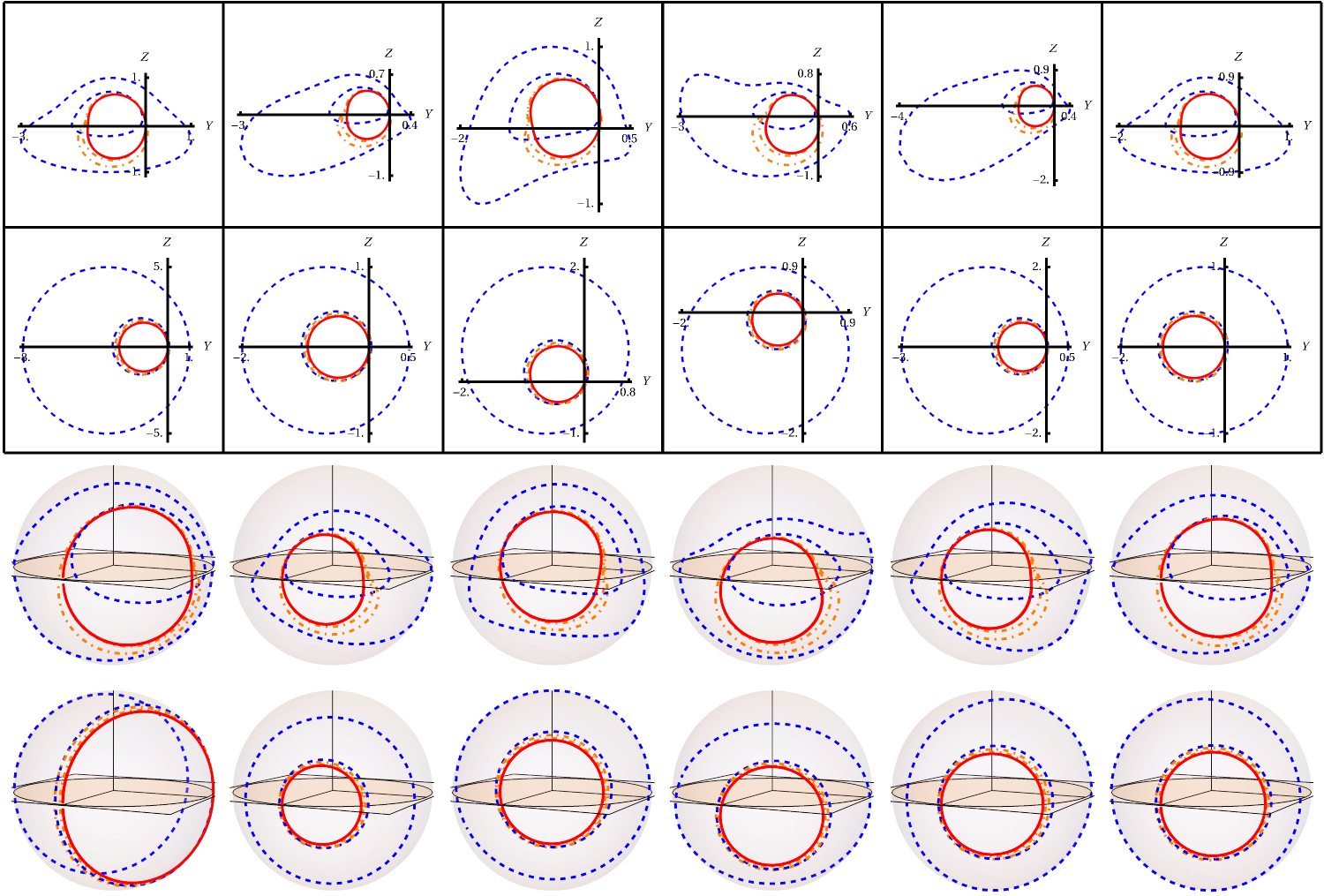}
  \caption{Primary (dashed), secondary (dashed-dotted), and $n$=2 images of the thin accretion disk around a Kerr-de Sitter black hole with the spin parameter of $a=0.99M$ and the cosmological constant of $\Lambda=0.01M^{-2}$ for moving observers with respect to the co-moving frame. Specifically, the moving observers are located at the distance of $r_{\rm o}=16M$ and the inclination angle of $\theta_{\rm o}=2\pi/5$ (top panels) and $\theta_{\rm o}=\pi/25$ (bottom panels) with 4-velocities $u^{(r, \pm, \rm c)}$, $u^{(\theta, \pm, \rm c)}$, and $u^{(\phi, \pm, \rm c)}$ at a speed of $\upsilon=0.5$, respectively. \label{F10}}
\end{figure}

\begin{figure}
  \includegraphics[width=1\linewidth]{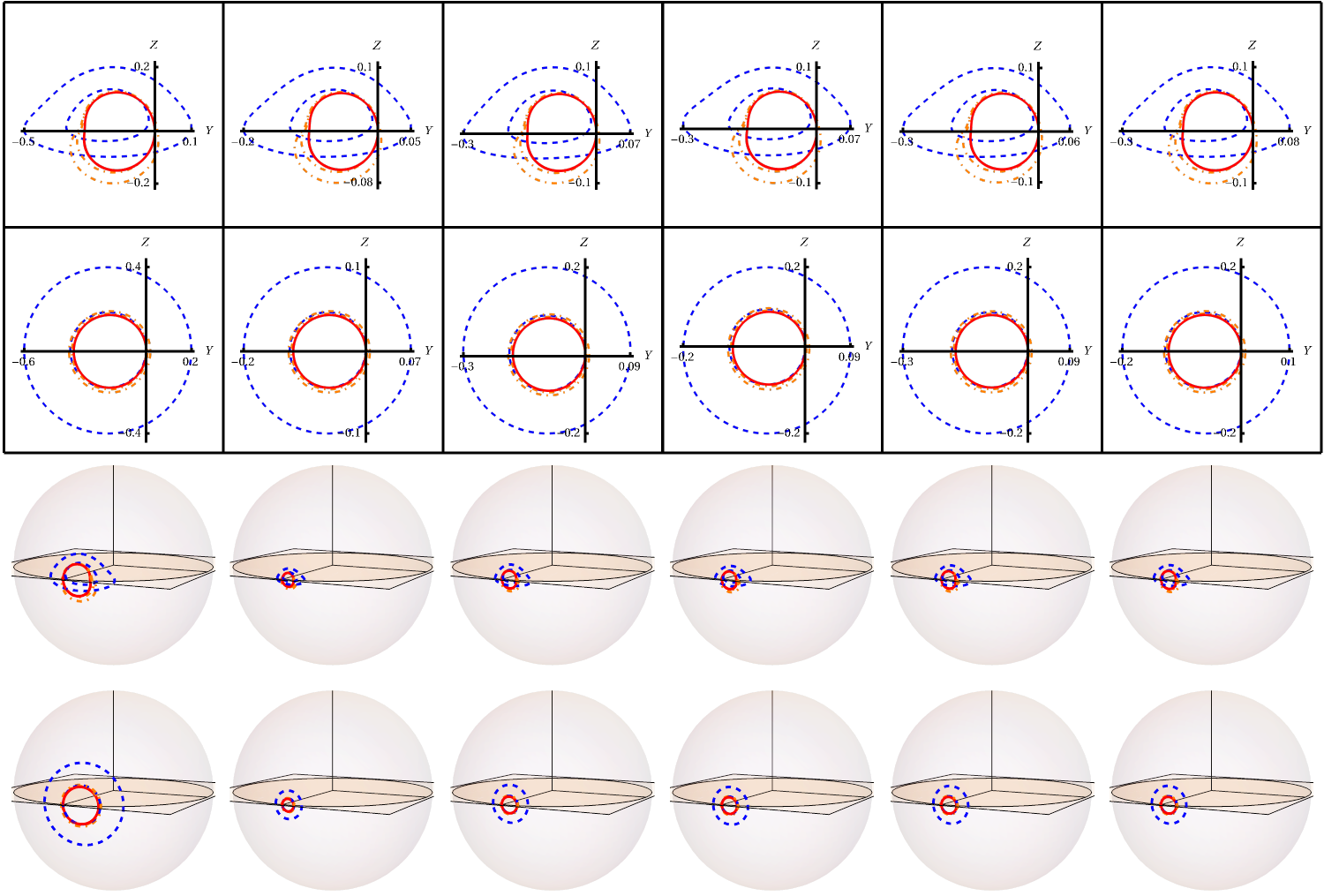}
  \caption{Primary (dashed), secondary (dashed-dotted), and $n$=2 images of the thin accretion disk around a Kerr-de Sitter black hole with the spin parameter of $a=0.99M$ and the cosmological constant of $\Lambda=0.00029M^{-2}$ for moving observers with respect to the co-moving frame. Specifically, the moving observers are located at the distance of $r_{\rm o}=100M$ and the inclination angle of $\theta_{\rm o}=2\pi/5$ (top panels) and $\theta_{\rm o}=\pi/25$ (bottom panels) with 4-velocities $u^{(r, \pm, \rm c)}$, $u^{(\theta, \pm, \rm c)}$, and $u^{(\phi, \pm, \rm c)}$ at a speed of $\upsilon=0.5$, respectively. \label{F11}}
\end{figure}

\begin{figure}
  \includegraphics[width=1\linewidth]{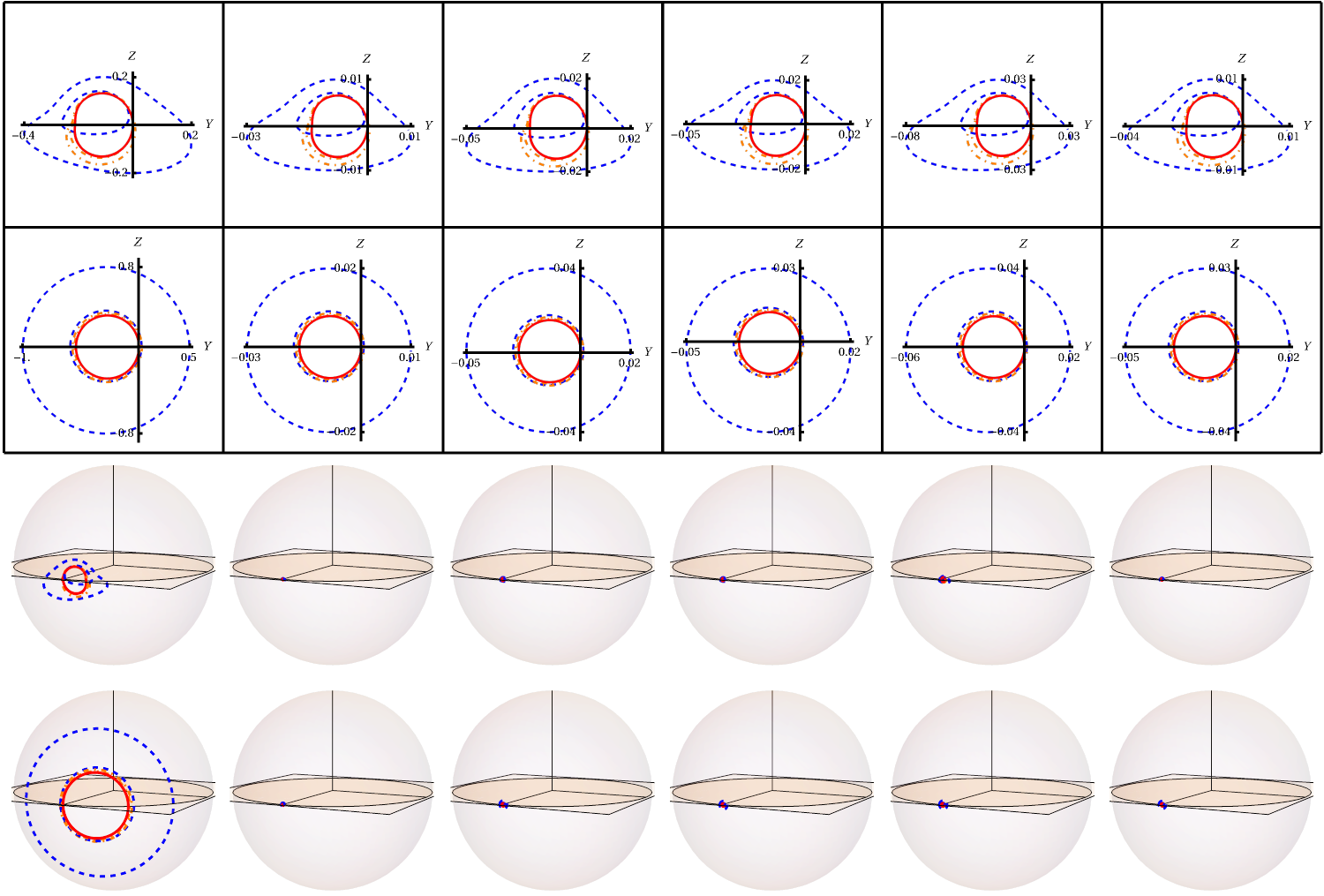}
  \caption{Primary (dashed), secondary (dashed-dotted), and $n$=2 images of the thin accretion disk around a Kerr-de Sitter black hole with the spin parameter of $a=0.99M$ and the cosmological constant of $\Lambda=0.01M^{-2}$ for moving observers with respect to the static frame. Specifically, the moving observers are located at the distance of $r_{\rm o}=16M$ and the inclination angle of $\theta_{\rm o}=2\pi/5$ (top panels) and $\theta_{\rm o}=\pi/25$ (bottom panels) with 4-velocities $u^{(r, \pm, \rm s)}$, $u^{(\theta, \pm, \rm s)}$, and $u^{(\phi, \pm, \rm s)}$ at a speed of $\upsilon=0.5$, respectively. \label{F12}}
\end{figure}

One might also be interested in the lower-order images for observers near the outer horizons. Figures~\ref{F10} and \ref{F11} show the lower-order images for observers with respect to the co-moving frame. It is evident that the degree of distortion is reduced as the distance increases, once again. There appears to be no additional effect for observers near the outer horizon. For comparison, we also consider the images for moving observers with respect to the static frame in Figure \ref{F12}. This reproduces the results in previous studies indicating that the size of the black hole image tends to vanish when approaching outer horizons \cite{Perlick:2018iye, Chang:2019vni}. For the cases shown in Figure~\ref{F10} and \ref{F12}, we let the moving observers have the same locations but be situated in different frames. It is found that the variation of the images is more sensitive to the motions of observers in the co-moving frame.

\subsection{Aberration formula \label{V.C}}

In the previous sections, we qualitatively studied the variation of lower-order images of the thin accretion disk for observers in motion. Here, we will present several quantitative results regarding the size of lower-order images.

\begin{figure}
  \includegraphics[width=1\linewidth]{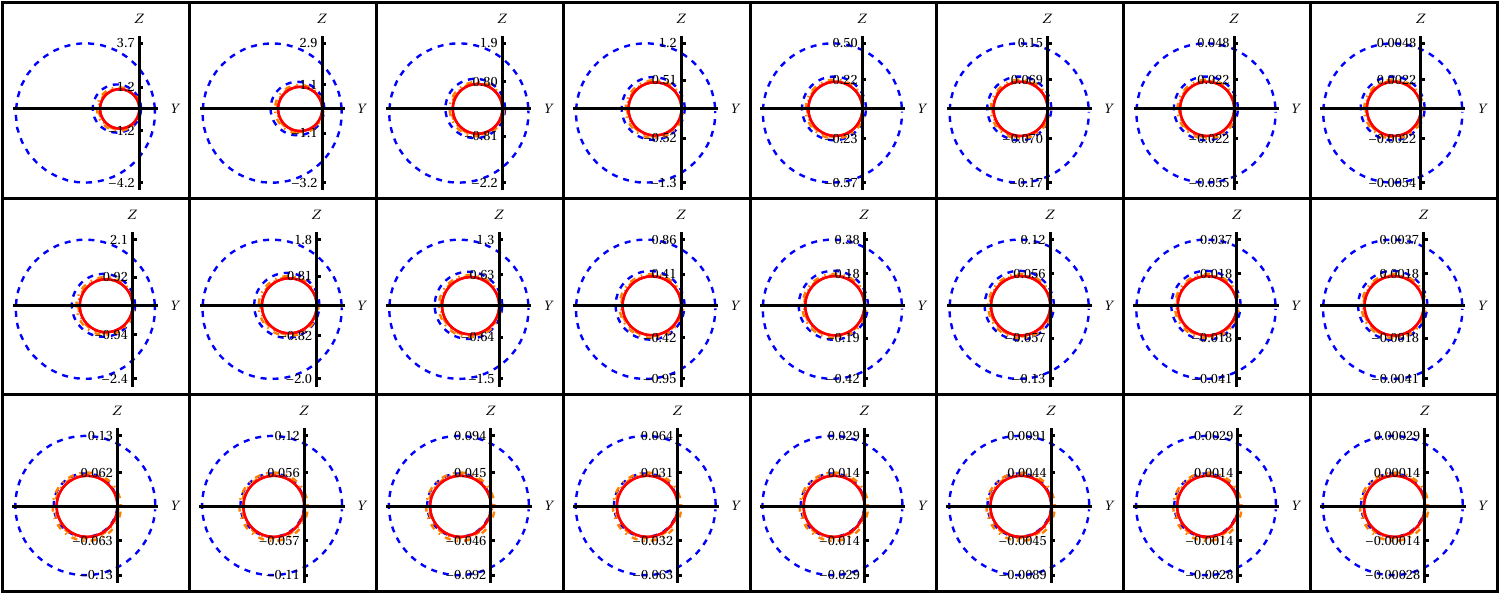}
  \caption{Primary (dashed), secondary (dashed-dotted), and $n$=2 images of the thin accretion disk around a Kerr-de Sitter black hole (top panel) or a Kerr black hole (center and bottom panels) with a spin parameter of $a=0.99M$ and a cosmological constant of $\Lambda=0.01M^{-2}$ for in-going observers. Specifically, the observers are located at $r_{\rm o}=10M$ (top and center panels) and $r_{\rm o}=100M$ (bottom panels) and an inclination angle of $\theta_{\rm o}=\pi/25$ and a 4-velocity $u^{(r,-,\rm c)}$ at speeds $\upsilon$ as $\upsilon$ as $0$, $0.1$, $0.3$, $0.6$, $0.9$, $0.99$, $0.999$, $0.99999$. \label{F13}}
\end{figure}

\begin{figure}
  \includegraphics[width=1\linewidth]{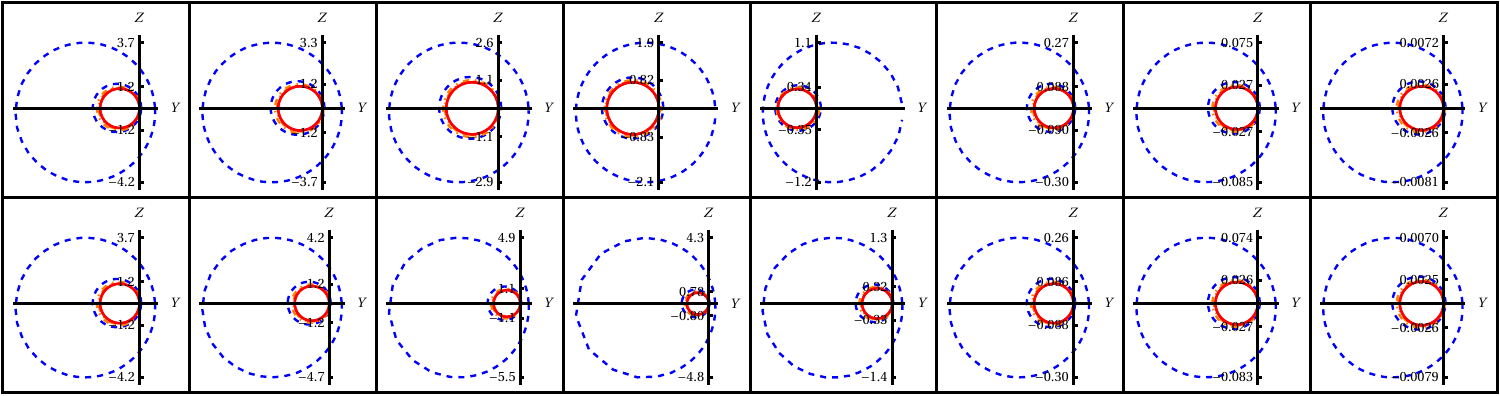}
  \caption{Primary (dashed), secondary (dashed-dotted) and $n$=2 images of the thin accretion disk around Kerr-de Sitter black hole with spin parameter $a=0.99M$ and cosmological constant $\Lambda=0.01M^{-2}$ for observers in axial $\phi$-motions. Specifically, the observers are located at $r_{\rm o} = 10 M$ and inclination angle $\theta_{\rm o}=\pi/25$  with 4-velocity  $u^{(\phi,-,\rm c)}$ (top pannels) and $u^{(\phi,+,\rm c)}$ (bottom panels) in the speeds $\upsilon$ as $0$, $0.1$, $0.3$, $0.6$, $0.9$, $0.99$, $0.999$, $0.99999$, respectively. \label{F14}}
\end{figure}

\begin{figure}
  \includegraphics[width=1\linewidth]{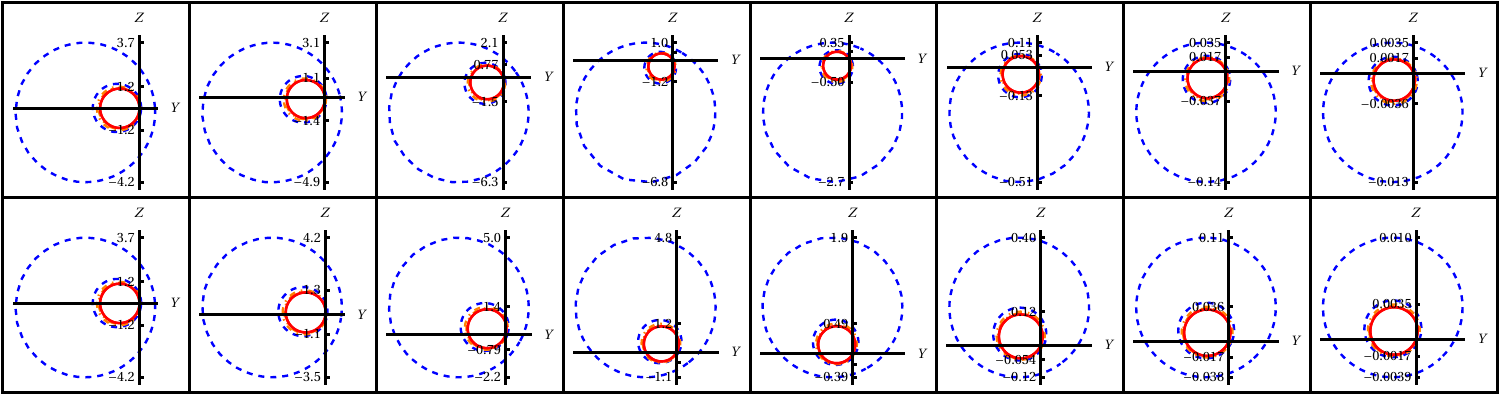}
  \caption{Primary (dashed), secondary (dashed-dotted) and $n$=2 images of the thin accretion disk around Kerr-de Sitter black hole with spin parameter $a=0.99M$ and cosmological constant $\Lambda=0.01M^{-2}$ for observers in axial $\theta$-motions. Specifically, the observers are located at $r_{\rm o} = 10 M$ and inclination angle $\theta_{\rm o}=\pi/25$  with 4-velocity  $u^{(\theta,-,\rm c)}$ (top pannels) and $u^{(\theta,+,\rm c)}$ (bottom panels) in the speeds $\upsilon$ as $0$, $0.1$, $0.3$, $0.6$, $0.9$, $0.99$, $0.999$, $0.99999$, respectively. \label{F15}}
\end{figure}

Figure~\ref{F13} displays the images for in-going observers at different speeds $\upsilon$. It is evident that the size of the images decreases with increasing $\upsilon$. Additionally, we also consider observers in axial motion in Figures~\ref{F14} and \ref{F15}. In addition to variations in the image size, we find that the relative distortion between primary, secondary, and $n=2$ images tends to remain fixed as $\upsilon$ approaches the speed of light.
To quantitively study the aberration effect, we introduce the relative size on the projection plane $(Y, Z)$, namely,
\begin{equation}
  Z_{\rm size} \equiv \frac{(Z_{\rm max} -Z_{\rm min})|_\upsilon}{(Z_{\rm max} -Z_{\rm min})|_{\upsilon=0}}~. \label{zsize}
\end{equation} 
On the other hand, the Penrose aberration formula takes the form of $\tan\Psi'=\tan\Psi \sqrt{(1-\upsilon)(1+\upsilon)}$ \cite{1959PCPS...55..137P}, where the angular diameter $\Psi'$ and $\Psi$ can be defined using celestial coordinates in different frames. Therefore, by providing a stereographic projection on the observer's celestial sphere, we can relate the quantity defined in Eq.~(\ref{zsize}) with Penrose's aberration formula due to $Z_{\rm size}=\tan\Psi'/\tan\Psi$.

\begin{figure}
  \includegraphics[width=.8\linewidth]{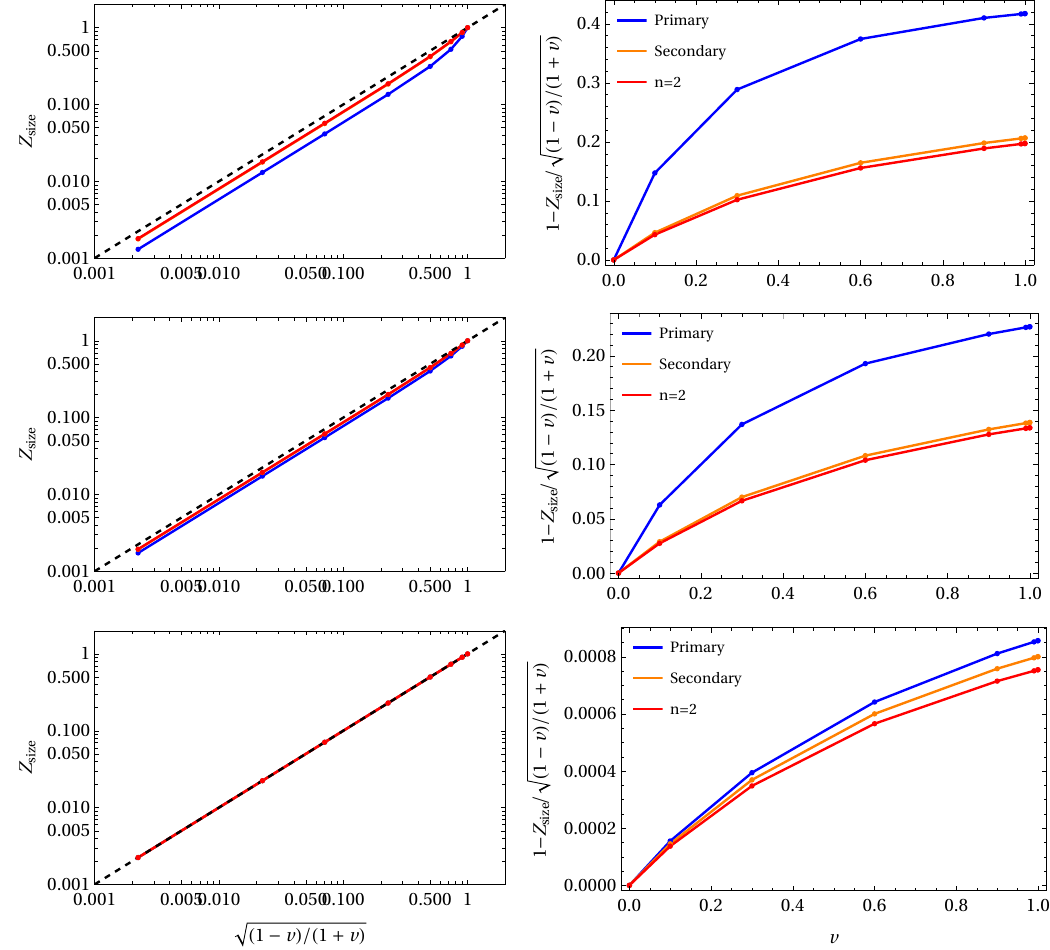}
  \caption{Left panel: the relative size of images as functions of the $\sqrt{(1-\upsilon)(1+\upsilon)}$. Right panel:  deviation of the relative size as functions of the $v$. These plots are based on the case given in Figure~\ref{F13}. \label{F16}}
\end{figure}

\begin{figure}
  \includegraphics[width=.8\linewidth]{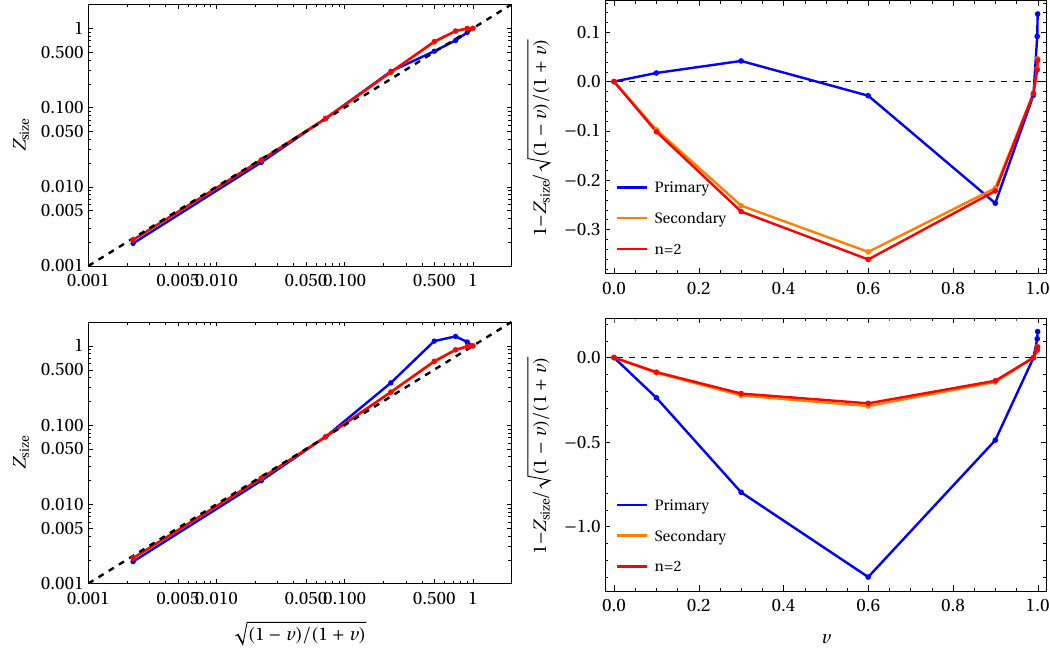}
  \caption{Left panel: the relative size of images as functions of the $\sqrt{(1-\upsilon)(1+\upsilon)}$. Right panel: deviation of the relative size as functions of the $v$. These plots are based on the case given in Figure~\ref{F14}. \label{F17}}
\end{figure}

\begin{figure}
  \includegraphics[width=.8\linewidth]{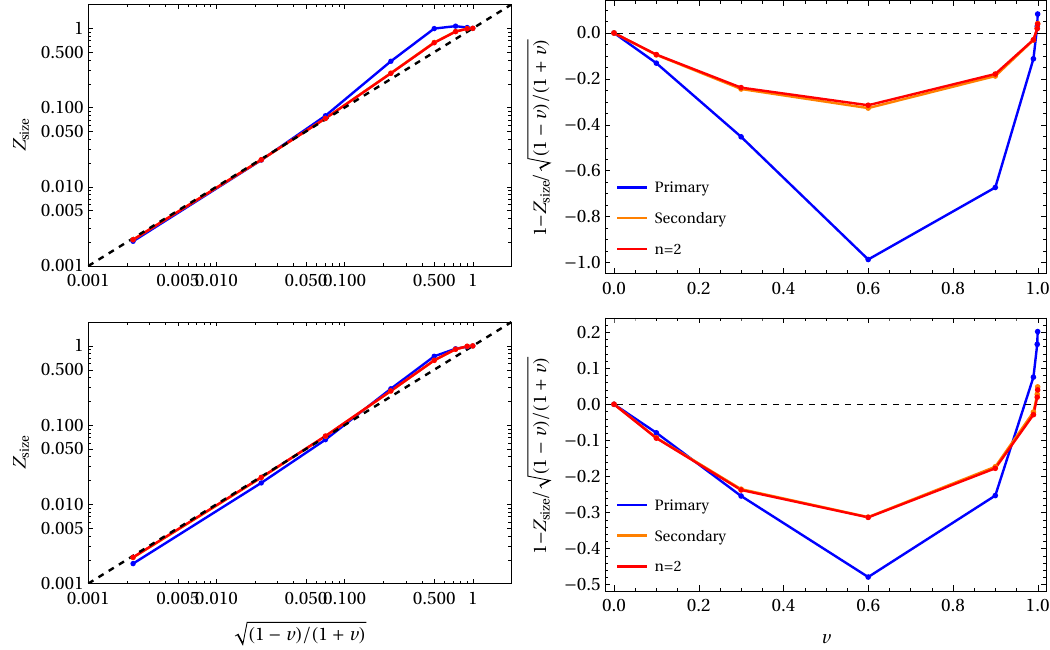}
  \caption{Left panel: the relative size of images as functions of the $\sqrt{(1-\upsilon)(1+\upsilon)}$. Right panel: deviation of the relative size as functions of the $v$. These plots are based on the case given in Figure~\ref{F15}. \label{F18}}
\end{figure}

In Figures~\ref{F16}--\ref{F18}, we present $Z_{\rm size}$ as functions of the speeds $\upsilon$ based on the cases given in Figures~\ref{F13}--\ref{F15}, respectively. It shows that the factor $\sqrt{(1-\upsilon)(1+\upsilon)}$ can overall describe the variation of the primary, secondary, or $n=2$ images. The consistency indicates that quantities $\upsilon$ can at least function as a relative 3-speed in physics. Based on $Z_{\rm size}/\sqrt{(1-\upsilon)(1+\upsilon)}$ in the right panels of Figures~\ref{F16}--\ref{F18}, there is a deviation from Penrose's aberration formula, which might be understood as the influence from the gravity environment of observers. For radial motions, the deviation is shown to be larger for near observers.
And the variation of the secondary and $n=2$ images is hardly distinguishable in the view of near observers compared to the case of distant observers.  We also quantitatively find the asymptotical behavior of the images, namely, the relative distortion between primary, secondary, and $n=2$ images tends to be fixed as $\upsilon$ approaches $1$. 
For axial motions shown in Figures~\ref{F17} and \ref{F18}, the deviation does not vary monotonically with the speed $\upsilon$, which is different from the cases of radial motions. Additionally, one might also find the frame-dragging effect by comparing the right panels of Figure~\ref{F17}.



\subsection{Upper bound of speed?}
\begin{figure}
  \includegraphics[width=0.8\linewidth]{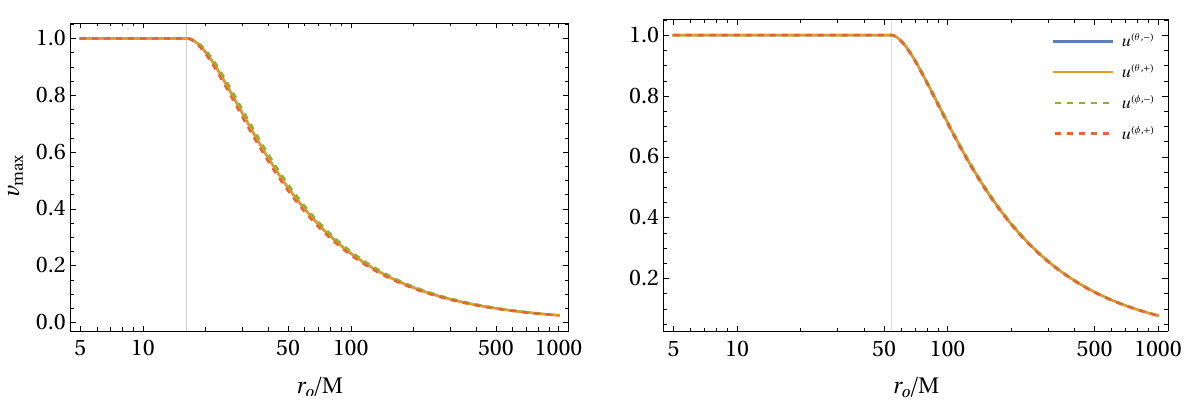}
  \caption{The maximum speeds of axial motions with respect to the co-moving frame as a function of the distance to the Kerr-de Sitter black hole. Left panel: the black hole parameter $a=0.99M$ and $\Lambda=0.01M^{-2}$. Right panel: the black hole parameter $a=0.99M$ and $\Lambda=0.001M^{-2}$. In both panels, observers are situated at the inclination angle $\theta_{\rm o}=2\pi/5$. \label{F19}}
\end{figure}

The above results suggest that the quantities $\upsilon$ can be used as a physical 3-speed for interpreting the aberration formula. However, as shown in Figure~\ref{F8}, it was found that there is an upper bound of $\upsilon$ for observers beyond the outer horizon in axial motions.  For illustration, we present the maximum speeds $\upsilon_{\rm max}$ as a function of the observers' distance to the Kerr-de Sitter black hole in Figure~\ref{F19}.
As the reference frame approaches the outer horizon, the maximum speed is recovered to the speed of light.

As mentioned before, the 4-velocities in Eqs.~(\ref{u:int:com}) are instantaneous velocities. For example, the $u^{(r,\pm,\rm c)}$ is obtained by substituting integral constants $\kappa(\bm x_{\rm o})$ and $\lambda(\bm x_{\rm o})$ in Eqs.~(\ref{u:gen}), such that $(u^{(r,\pm,\rm c),\theta}-u^{\theta}_{\rm c})|_{{\bm x}_{\rm o}} = (u^{(r,\pm,\rm c),\phi}-u^{\phi}_{\rm c})|_{{\bm x}_{\rm o}} = 0$. Thus, Eqs.~(\ref{u:int:com}) merely provide the expressions of the 4-velocities at $\bm x_{\rm o}$. There seems to be no problem when considering the observers within the outer horizon. The naive approach seems sufficient for obtaining representative 4-velocities in various directions. However, we still believe that the upper bound of speed has no clear physical origin due to the coordinate-dependent nature of the above approach. We can not exclude the possibility that the 4-velocities are ill-defined here when they are beyond the outer horizon of the Kerr-de Sitter black hole.

The thought experiment concerning the launch of telescopes, as discussed in Section~\ref{V.B}, should be distinguished from the procedure for obtaining 4-velocities relative to the co-moving frame. The latter only involves technical aspects. Since $\upsilon$ is a coordinate-independent quantity, the method used to determine axial 4-velocities may not significantly impact the outcomes of studying the variations of lower-order images. Obtaining a proper axial 4-velocity may be beyond the scope of this study and would be addressed in future research.

\

\section{Conclusions and discussions \label{VI}}

This paper studied the primary, secondary and $n=2$ images of the thin accretion disk in the view of finite distant observers in motions with the astrometric approach. Specifically, we considered the static, and co-moving observers, as well as observers in radial and axial motions with respect to the static, or co-moving frame. 
The study revealed that the shapes of lower-order images get distorted for observers in motion. Notably, the variation in primary images was shown to be more sensitive than that of the $n=2$ images. We also quantitively studied the aberration effect on the size of the lower-order images and compared it with Penrose's aberration formula. Although the aberration formula can describe the variation of the image sizes, overall, one can still find slight deviations for primary, secondary and $n=2$ images. Specifically, the behaviors of primary images exhibit the most pronounced deviations from the aberration formula. 
We anticipate that the distinct behaviors of different-order images might offer a novel approach to investigating both space-time geometries and emissions separately.

On technique aspects, the astrometric approach provides a coordinate-independent and tetrad-independent method for establishing observers' celestial spheres. In previous studies, it was solely employed for calculating the black hole shadow \cite{Chang:2020miq, Chang:2020lmg, Chang:2021ngy, He:2020dfo}. Here, we expanded the framework to include lower-order images of emissions. Additionally, we introduced an alternative form of transfer functions for the analytical ray-tracing scenario pioneered by others \cite{Gralla:2019ceu, Gralla:2019drh, Cardenas-Avendano:2022csp}. This alternative formulation showed that the $n$th-order images of the thin accretion disk can be obtained by substituting the new variable $|\chi_{\rm s}|=\pi(n+1/2)$ into our transfer functions.


We presented numerous intuitive results in Section~\ref{V}. However, in Section~\ref{V.B}, the exploration of the distorted shapes of the primary, secondary, and $n=2$ images is primarily qualitative due to the absence of a robust quantity formulating the shapes. Perhaps, further studies may delve into addressing this aspect.

\smallskip
{\it Acknowledgments. }
This work will be supported by the National Nature Science Foundation of China under grant No.~12305073.

\appendix
\section{Gallery of intensity images}
To show the robustness of our results, we present the intensity images of the thin accretion disk regarding the observers in motions. Theoretically, it has no relevance to the aberration effect we have studied in this paper. 

\begin{figure} 
  \includegraphics[width=1\linewidth]{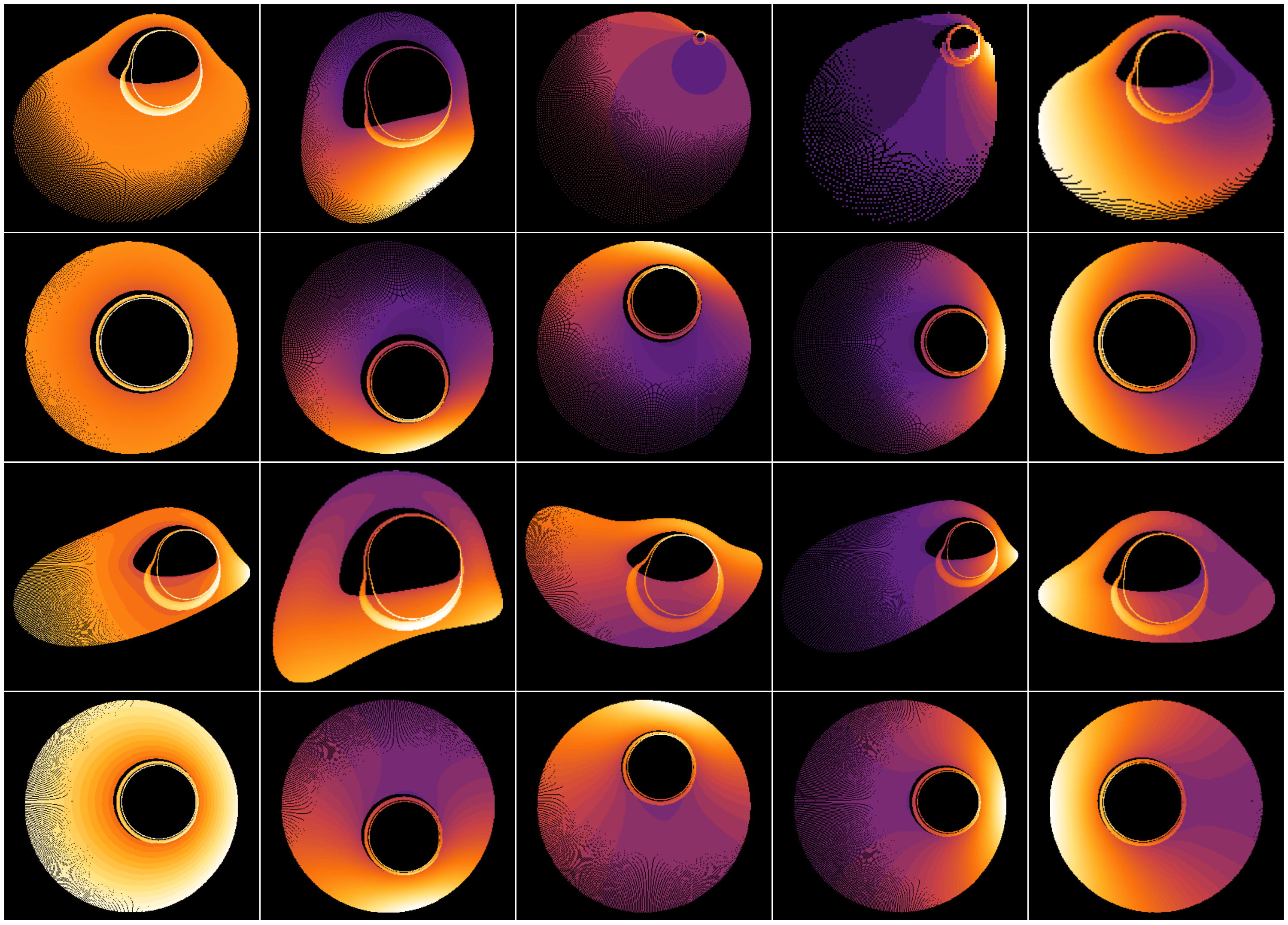} 
  \caption{Intensity images of thin accretion disk for Kerr(-de Sitter) black hole with respect to near observers. \label{F20}}
\end{figure}

\begin{figure}
  \includegraphics[width=1\linewidth]{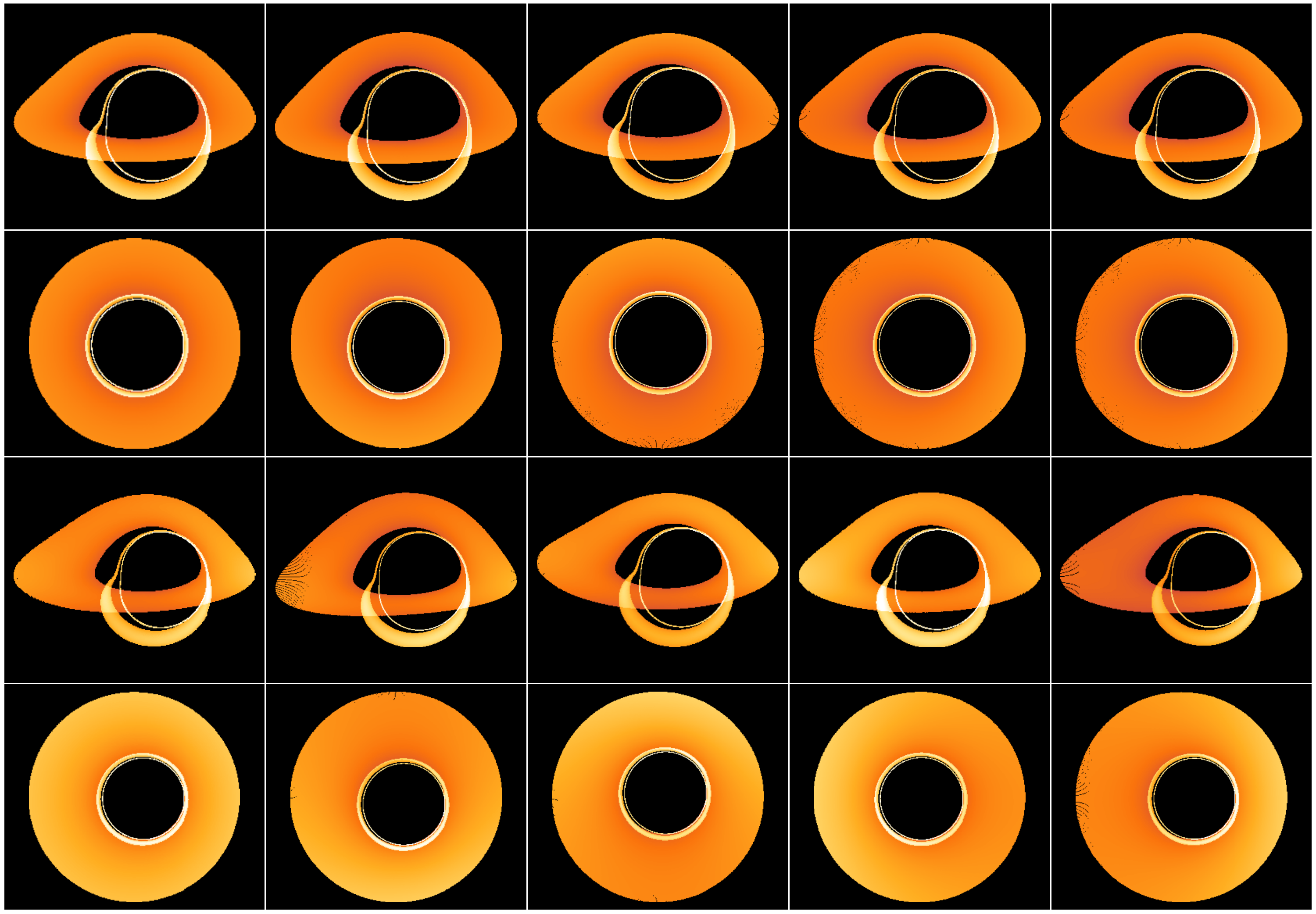}
  \caption{Intensity images of thin accretion disk for Kerr(-de Sitter) black hole with respect to distant observers. \label{F21}}
\end{figure}

In Figures~\ref{F20} and \ref{F21}, we let the distribution of emissions intensity in Eq.~(\ref{Iemt}) be a constant function, namely, $f_{\rm d}(r)=1$. It indicates $I_{\rm emt}(\bm x)=1$, $2$, $3$ for primary, secondary and $n=2$ images, respectively. Here, the gradual variations in colors are determined by the redshift formulated in Eq.~(\ref{red}). One should be cautious, as there could be unphysical image distortion in the images in the view of the near observers shown in Figure~\ref{F20}. In this case, one can reference the reliable results shown in the bottom panels of Figure~\ref{F5} and \ref{F10} on the celestial sphere.

\bibliography{ref}
\end{document}